%% file: IRAS4A-outflows-Chahine.tex
\documentclass[fleqn,usenatbib]{mnras}

\usepackage[T1]{fontenc}
\usepackage{wrapfig}
\usepackage{placeins}

\DeclareRobustCommand{\VAN}[3]{#2}
\let\VANthebibliography\thebibliography
\def\thebibliography{\DeclareRobustCommand{\VAN}[3]{##3}\VANthebibliography}

\usepackage{graphicx}	
\usepackage{amsmath}	
\usepackage{multirow}
\usepackage{lscape}

\usepackage{newtxtext,newtxmath}
\usepackage{makecell}
\usepackage[super]{nth}

\title[IRAS\,4A outflows]{Multiple chemical tracers finally unveil the intricate NGC\,1333 IRAS\,4A outflow system. FAUST XVI}

\author[L. Chahine et al.]{Layal Chahine,$^{1}$\thanks{E-mail: layal.chahine@univ-grenoble-alpes.fr}
    Cecilia Ceccarelli,$^{1}$\thanks{E-mail: cecilia.ceccarelli@univ-grenoble-alpes.fr}
    Marta De Simone,$^{2,1,3}$
    Claire J. Chandler,$^{4}$
    Claudio Codella,$^{3}$
    \newauthor
    Linda Podio,$^{3}$
    Ana L\'{o}pez-Sepulcre,$^{1,5}$
    Nami Sakai,$^{6}$
    Laurent Loinard,$^{7,8}$
    Mathilde Bouvier,$^{9}$
    Paola Caselli,$^{10}$
    \newauthor
    Charlotte Vastel,$^{11}$
    Eleonora Bianchi,$^{12,1}$
    Nicol\'as Cuello,$^{1}$
    Francesco Fontani,$^{3,10,13}$
    Doug Johnstone,$^{14,15}$
    \newauthor 
    Giovanni Sabatini,$^{3}$
    Tomoyuki Hanawa,$^{16}$
    Ziwei E. Zhang,$^{6}$
    Yuri Aikawa,$^{17}$
    Gemma Busquet,$^{18}$
    Emmanuel Caux,$^{11}$
    \newauthor
    Aurore Dur\'{a}n,$^{7}$
    Eric Herbst,$^{19}$
    Fran\c{c}ois M\'enard,$^{1}$
    Dominique Segura-Cox,$^{10,20}$
    Brian Svodoba,$^{4,21}$
    \newauthor
    Nadia Balucani,$^{22}$
    Steven Charnley,$^{23}$
    Fran\c{c}ois Dulieu,$^{24}$
    Lucy Evans,$^{25}$
    Davide Fedele,$^{3}$
    Siyi Feng,$^{26}$
    \newauthor
    Tetsuya Hama,$^{27,28}$
    Tomoya Hirota,$^{29}$
    Andrea Isella,$^{30}$
    Izaskun J\'{i}menez-Serra,$^{31}$
    Bertrand Lefloch,$^{32}$
    \newauthor
    Luke T. Maud,$^{2}$
    Mar\'{i}a Jos\'{e} Maureira,$^{10}$
    Anna Miotello,$^{2}$
    George Moellenbrock,$^{4}$
    Hideko Nomura,$^{33}$
    \newauthor
    Yasuhiro Oba,$^{34}$
    Satoshi Ohashi,$^{6}$
    Yuki Okoda,$^{35,6}$
    Yoko Oya,$^{36,35}$
    Jaime Pineda,$^{10}$
    Albert Rimola,$^{37}$
    \newauthor
    Takeshi Sakai,$^{38}$
    Yancy Shirley,$^{39}$
    Leonardo Testi,$^{40,3}$
    Serena Viti,$^{9}$
    Naoki Watanabe,$^{34}$
    Yoshimasa Watanabe,$^{41}$
    \newauthor
    Yichen Zhang,$^{6}$
    and Satoshi Yamamoto$^{42}$
    \\\\
$^{1}$Univ. Grenoble Alpes, CNRS, IPAG, 38000 Grenoble, France\\
$^{2}$ESO, Karl Schwarzchild Srt. 2, 85478 Garching bei Munchen, Germany \\
$^{3}$INAF, Osservatorio Astrofisico di Arcetri, Largo E. Fermi 5, I-50125, Firenze, Italy\\
$^{4}$National Radio Astronomy Observatory, PO Box O, Socorro, NM 87801, USA \\
$^{5}$Institut de Radioastronomie Millim\'etrique (IRAM), 300 rue de la Piscine, 38406 Saint-Martin-d'H\`eres, France \\
$^{6}$RIKEN Cluster for Pioneering Research, 2-1, Hirosawa, Wako-shi, Saitama 351-0198, Japan \\
$^{7}$Instituto de Radioastronom\'{i}a y Astrof\'{i}sica, Universidad Nacional Aut\'{o}noma de M\'{e}xico, A.P. 3-72 (Xangari), 8701, Morelia, Mexico \\
$^{8}$Instituto de Astronom\'{i}a, Universidad Nacional Aut\'{o}noma de M\'{e}xico, Ciudad Universitaria, A.P. 70-264, Cuidad de M\'{e}xico 04510, Mexico \\
$^{9}$Leiden Observatory, Leiden University, P.O. Box 9513, 2300 RA Leiden, The Netherlands \\
$^{10}$Center for Astrochemical Studies, Max-Planck-Institut f\"{u}r extraterrestrische Physik (MPE), Gie{\ss}enbachstr. 1, D-85741 Garching, Germany \\
$^{11}$IRAP, Universit\`{e} de Toulouse, CNRS, CNES, UPS, Toulouse, France \\
$^{12}$ORIGINS, Excellence Cluster Origins, Boltzmannstrasse 2, D-85748 Garching bei München, Germany\\
$^{13}$LERMA, Observatoire de Paris, PSL Research University, CNRS, Sorbonne Universit\'{e}, 92190 Meudon, France\\
$^{14}$NRC Herzberg Astronomy and Astrophysics, 5071 West Saanich Road, Victoria, BC, V9E 2E7, Canada \\
$^{15}$Department of Physics and Astronomy, University of Victoria, Victoria, BC, V8P 5C2, Canada\\
$^{16}$Center for Frontier Science, Chiba University, 1-33 Yayoi-cho, Inage-ku, Chiba 263-8522, Japan\\
$^{17}$Department of Astronomy, The University of Tokyo, 7-3-1 Hongo, Bunkyo-ku, Tokyo 113-0033, Japan \\
$^{18}$Departament de F\'isica Qu\`antica i Astrof\'isica, Institut de Ci\`encies del Cosmos, Universitat de Barcelona (IEEC-UB),  Mart\'i Franqu\`es, 1, 08028 Barcelona, Spain \\
$^{19}$Department of Chemistry, University of Virginia, McCormick Road, PO Box 400319, Charlottesville, VA 22904, USA\\
$^{20}$The University of Texas at Austin, 2515 Speedway, Austin, Texas 78712, USA\\
$^{21}$Jansky Fellow of the National Radio Astronomy Observatory\\
$^{22}$Department of Chemistry, Biology, and Biotechnology, The University of Perugia, Via Elce di Sotto 8, 06123 Perugia, Italy\\
$^{23}$Astrochemistry Laboratory, Code 691, NASA Goddard Space Flight Center, 8800 Greenbelt Road, Greenbelt, MD 20771, USA\\
$^{24}$CY Cergy Paris Universit\'{e}, Sorbonne Universit\'{e}, Observatoire de Paris, PSL University, CNRS, LERMA, F-95000, Cergy, France\\
$^{25}$School of Physics and Astronomy, University of Leeds, Leeds LS2 9JT, UK\\
$^{26}$Department of Astronomy, Xiamen University, Xiamen, Fujian 361005, P. R. China\\
$^{27}$Komaba Institute for Science, The University of Tokyo, 3-8-1 Komaba, Meguro, Tokyo 153-8902, Japan\\
$^{28}$Department of Basic Science, The University of Tokyo, 3-8-1 Komaba, Meguro, Tokyo 153-8902, Japan\\
$^{29}$National Astronomical Observatory of Japan, Osawa 2-21-1, Mitaka-shi, Tokyo 181-8588, Japan\\
$^{30}$Department of Physics and Astronomy, Rice University, 6100 Main Street, MS-108, Houston, TX 77005, USA\\
$^{31}$Centro de Astrobiolog\'{\i}a (CSIC-INTA), Ctra. de Torrej\'on a Ajalvir, km 4, 28850, Torrej\'on de Ardoz, Spain\\
$^{32}$Université de Bordeaux – CNRS Laboratoire d’Astrophysique de Bordeaux, 33600 Pessac, France\\
$^{33}$Division of Science, National Astronomical Observatory of Japan, 2-21-1 Osawa, Mitaka, Tokyo 181-8588, Japan\\
$^{34}$Institute of Low Temperature Science, Hokkaido University, N19W8, Kita-ku, Sapporo, Hokkaido 060-0819, Japan\\
$^{35}$Department of Physics, The University of Tokyo, 7-3-1, Hongo, Bunkyo-ku, Tokyo 113-0033, Japan\\
$^{36}$Yukawa Institute for Theoretical Physics, Kyoto University Oiwake-cho, Kitashirakawa, Sakyo-ku, Kyoto-shi, Kyoto-fu 606-8502, Japan\\
$^{37}$Departament de Qu\'{i}mica, Universitat Aut$\grave{o}$noma de Barcelona, 08193 Bellaterra, Spain\\
$^{38}$Graduate School of Informatics and Engineering, The University of Electro-Communications, Chofu, Tokyo 182-8585, Japan\\
$^{39}$Steward Observatory, 933 N Cherry Ave., Tucson, AZ 85721 USA\\
$^{40}$Dipartimento di Fisica e Astronomia “Augusto Righi” Viale Berti Pichat 6/2, Bologna, Italy\\
$^{41}$Materials Science and Engineering, College of Engineering, Shibaura Institute of Technology, 3-7-5 Toyosu, Koto-ku, Tokyo 135-8548, Japan\\
$^{42}$SOKENDAI, Shonan Village, Hayama, Kanagawa 240-0193, Japan}

\date{Accepted 2024 May 20. Received 2024 May 20; in original form 2024 February 29}

\pubyear{2024}

\begin{document}
\label{firstpage}
\pagerange{\pageref{firstpage}--\pageref{lastpage}}
\maketitle

\begin{abstract}

The exploration of outflows in protobinary systems presents a challenging yet crucial endeavour, offering valuable insights into the dynamic interplay between protostars and their evolution. In this study, we examine the morphology and dynamics of jets and outflows within the IRAS\,4A protobinary system. This analysis is based on ALMA observations of SiO(5--4), H$_2$CO(3$_{0,3}$--2$_{0,3}$), and HDCO(4$_{1,4}$--3$_{1,3}$) with a spatial resolution of $\sim$150\,au. Leveraging an astrochemical approach involving the use of diverse tracers beyond traditional ones has enabled the identification of novel features and a comprehensive understanding of the broader outflow dynamics. Our analysis reveals the presence of two jets in the redshifted emission, emanating from IRAS\,4A1 and IRAS\,4A2, respectively. Furthermore, we identify four distinct outflows in the region for the first time, with each protostar, 4A1 and 4A2, contributing to two of them. We characterise the morphology and orientation of each outflow, challenging previous suggestions of bends in their trajectories. The outflow cavities of IRAS\,4A1 exhibit extensions of 10$''$  and 13$''$  with position angles (PA) of 0$^{\circ}$ and -12$^{\circ}$, respectively, while those of IRAS\,4A2 are more extended, spanning 18$''$  and 25$''$ with PAs of 29$^{\circ}$ and 26$^{\circ}$. We propose that the misalignment of the cavities is due to a jet precession in each protostar, a notion supported by the observation that the more extended cavities of the same source exhibit lower velocities, indicating they may stem from older ejection events. 
\end{abstract}




\section{Introduction}
\input{sections/intro}

\section{Observations}
\input{sections/obs}

\section{Results}
\input{sections/results}


\section{Discussion}
\input{sections/discussion}


\section{Conclusions}
\input{sections/conclusion}

\section*{Acknowledgements}
We gratefully acknowledge Dr. Fabien Louvet for their valuable contribution to the discussion. 
This project has received funding from the EC H2020 research and innovation programme for the project ‘Astro-Chemical Origins’ (ACO; No. 811312) and the European Research Council (ERC) projects ‘The Dawn of Organic Chemistry’ (DOC; No.741002), and ‘Stellar-MADE’ (No.101042275). 
This paper makes use of the following ALMA data set: ADS/JAO.ALMA\#2018.1.01205.L (PI: Satoshi Yamamoto). ALMA is a partnership of the ESO (representing its member states), the NSF (USA) and NINS (Japan), together with the NRC (Canada) and the NSC and ASIAA (Taiwan), in cooperation with the Republic of Chile. The Joint ALMA Observatory is operated by the ESO, the AUI/NRAO, and the NAOJ. 
The authors thank the ALMA and NRAO staff for their support. 
Part of the data reduction/combination presented in this paper was performed using the GRICAD infrastructure (\url{https://gricad.univ-grenoble-alpes.fr}).
ClCo, LP, and GS acknowledge the PRIN-MUR 2020  BEYOND-2p (Astrochemistry beyond the second period elements, Prot. 2020AFB3FX), the PRIN MUR 2022 FOSSILS (Chemical origins: linking the fossil composition of the Solar System with the chemistry of protoplanetary disks, Prot. 2022JC2Y93), the project ASI-Astrobiologia 2023 MIGLIORA (Modeling Chemical Complexity, F83C23000800005), and the INAF-GO 2023 fundings PROTO-SKA (Exploiting ALMA data to study planet forming disks: preparing the advent of SKA, C13C23000770005). LP acknowledges the INAF Mini-Grant 2022 “Chemical Origins” (PI: L. Podio). GS acknowledges the INAF-Minigrant 2023 TRIESTE (“TRacing the chemIcal hEritage of our originS: from proTostars to planEts”; PI: G. Sabatini). ClCo and LP also acknowledge the financial support under the National Recovery and Resilience Plan (NRRP), Mission 4, Component 2, Investment 1.1, Call for tender No. 104 published on 2.2.2022 by the Italian Ministry of University and Research (MUR), funded by the European Union – NextGenerationEU– Project Title 2022JC2Y93 ChemicalOrigins: linking the fossil composition of the Solar System with the chemistry of protoplanetary disks – CUP J53D23001600006 - Grant Assignment Decree No. 962 adopted on 30.06.2023 by the Italian Ministry of Ministry of University and Research (MUR). EB acknowledges support from the Deutsche Forschungsgemeinschaft (DFG, German Research Foundation) under Germany´s Excellence Strategy – EXC 2094 – 390783311. M.B. acknowledges support from the European Research Council (ERC) Advanced Grant MOPPEX 833460. SBC was supported by the NASA Planetary Science Division Internal Scientist Funding Program through the Fundamental Laboratory Research work package (FLaRe). 





\section*{Data Availability}

The raw data are available on the ALMA archive at the end of the proprietary period (ADS/JAO.ALMA\#2018.1.01205.L). 
 



\bibliographystyle{mnras}
\bibliography{references} 



\clearpage
\appendix
\newpage

\FloatBarrier
\section{Maps}

\input{sections/apendix}


\bsp	
\label{lastpage}
\end{document}

%% file: sections/intro.tex
\label{sec:introduction}
It has long been known that the gravitational collapse, at the basis of the formation of a star and planetary system, is inevitably accompanied by a simultaneous ejection process, mediated by the magnetic field \citep[e.g.][]{Frank2014-pp6, Bally2014-PP6, Lee2020AARv}.
These ejections form the so-called "outflow systems", in general, constituted by supersonic jets, bow-shocks and cavities, all more or less delineated and identified by shocked material \citep[see, e.g., Fig. 3 of][]{Bally2016-ARAA, Rivera-Ortiz2023}.
Several molecular lines are generally used for these analyses.
Particularly useful are lines from species that have a very low abundance in molecular clouds because they are trapped in the interstellar grains, both in the refractory cores and volatile mantles.
The paradigmatic species is SiO, considered a clear-cut shock hallmark \citep[e.g.][]{Schilke1997, Caselli1997, Gusdorf2008b}.
Formaldehyde is also a good tracer of material ejected from the grain mantles, while deuterated formaldehyde traces the most recent ejected material \citep[e.g.][]{Fontani2014-Dformaldehyde}.

Often, but not always, the outflows have a bipolar structure.
Outflows have been the focus of a multitude of studies at the beginning because they were unexpectedly observed towards protostars before any sign of collapse was found \citep[e.g.][]{Kwan1976, Bally1983, 
Frank2014-pp6, Bally2016-ARAA}.
Soon after, it was realised the huge importance of the outflows in getting rid of the angular momentum of the protostellar infalling matter, as well as in shaping and disrupting the protostellar envelope and molecular cloud hosting the protostar.
Finally, molecular outflows cause different shocks, which in one way or another increase the gas chemical complexity \citep[e.g.][]{Bachiller1996-ARA&A} and are therefore very precious chemical laboratories \citep[e.g.][]{Codella2017-formamide}.

In addition to all the above reasons, the study of outflows in protobinary systems provides a fascinating glimpse into the complexity of star birth.
It sheds light on the genesis and early dynamical interactions of multiple protostars, providing essential insights into their intertwined evolution.
This is particularly true for systems where the binary companions are close enough to possibly interact and/or influence each other.
In fact, how the presence of a companion affects the ejection phenomenon is still not elucidated, especially in very young protobinary systems, where the surrounding envelope can mask and likely influence the evolution of the outflow itself.
It is not by chance that the best-studied systems, especially when it comes to chemistry, refer to single outflows (e.g. HH\,212; \citealt{Codella2007, Lee2015}, L1157-mm \citealt{Bachiller1995-methanol, Arce2013, Lefloch2017, Codella2017-formamide} and HH\,30 \citealt{Pety2006, Louvet2018} among others). 

The advent of high-angular resolution instruments has significantly advanced these studies, improving observational capacities and allowing for the resolution of smaller-scale features. 
Outflows in various multiple systems have now been studied, including but not limited to BHR\,71 \citep{Tobin2019}, IRAS\,16293-2422 \citep{Oya2021}, VLA\,1623–2417 \citep{Hsieh2020, Hara2021, Ohashi2022}, Ser-emb\,15 \citep{Sato2023}, OMC-2\,FIR\,4 \citep{Chahine2022b, Lattanzi2023, Sato2023-orion}, BHB2007 \citep{Hara2013, Alves2019, Zurlo2021, Evans2023}, HH\,1/2 (e.g. \citealt{Reipurth2000-hh1, Noriega-Crespo2012, Raga2015}) and the HH\,24 complex (e.g. \citealt{Reipurth2023}). 

A specific system that has piqued the interest of many researchers due to its distinctive features and complex dynamics and chemistry is IRAS\,4A. IRAS\,4A is a binary system, harbouring two protostars IRAS\,4A1 and IRAS\,4A2 (hereafter 4A1 and 4A2 respectively), located in the star-forming region NGC 1333 within the Perseus molecular cloud, at a distance of 293\,$\pm$\,22\,pc \citep{Zucker2018}. Notably, the system stands out for its remarkable outflow phenomena, which have been the subject of numerous studies. Indeed, the outflows within the IRAS\,4A system have been mapped using various molecular tracers, such as CO (e.g. \citealt{Blake1995, Girart1999, jorgensen2007, yildiz2012, Santangelo2015, Ching2016, Podio2021, Chuang2021}), CS \citep{DiFrancesco2001, Taquet2020}, SiO (e.g. \citealt{Lefloch1998, Choi2001, Santangelo2015, Koumpia2017, DeSimone2022}), SO and other sulphur-bearing species (e.g. \citealt{Santangelo2015, Taquet2020, Podio2021, Chuang2021}), as well as HCN (e.g. \citealt{Choi2001}), and H$_{2}$CO (e.g. \citealt{DiFrancesco2001, Su2019}). The presence of more complex molecules (CH$_{3}$OH and CH$_{3}$CHO; \citealt{DeSimone2022}) was revealed along the outflows, highlighting the rich chemistry of this region. 

The SiO observations at an angular resolution of 2$''$ \citep{Choi2005, Choi2006} suggested the presence of a bipolar northeast-southwest outflow driven by 4A2, and a tentative south outflow possibly driven by 4A1, with no distinct northern counterpart being evident. To determine the outflow direction, \cite{Choi2005} employed a least-squares fit method based on the positions of compact peaks near the driving source, identifying a trajectory passing near 4A2. Similar results were also found with other tracers. Using a higher angular resolution $\sim$1$''$, \cite{Santangelo2015} successfully distinguished the two outflows in the southern blue-shifted emission and, notably, at extremely high velocities (41 -- 55 km/s), they could resolve the redshifted counterpart of the 4A1 southern outflow for the first time. The same feature was also observed at similar velocity ranges with other tracers. There is no agreement, however, over the direction of the redshifted counterpart from 4A1. While the SiO emission mapped by \cite{Santangelo2015} suggests that the jet is oriented north-south (N-S), \cite{Ching2016} considered a position angle (PA) of -9$^{\circ}$, aligning with the trajectory of the southern counterpart of the jet. Nonetheless, \cite{Ching2016} have suggested that the northern outflow of IRAS\,4A1 is later bent due to the interaction with the outflow of IRAS\,4A2. A recent study using CO and SO at an angular resolution of 0.3$''$ suggested the presence of two northern outflow lobes, one from each protostar, both bent towards the northeast \citep{Chuang2021}. Another study at similar angular resolution using H$_{2}$CO suggested that both outflows are oriented north-south \citep{Su2019}. Meanwhile, the morphology of the southern blueshifted lobes is similar to that seen in previous studies, with the IRAS\,4A1 lobe oriented southeast and the IRAS\,4A2 lobe directed southwest. It is also worth noting that an abrupt bend at 20$''$ from 4A1 was also observed in several studies (e.g. \citealt{Choi2005, yildiz2012, DeSimone2022}). \cite{Choi2005} referred to this as “directional variability” and proposed that this feature may be attributed to external medium perturbations, specifically a deflection resulting from interactions with an external core or cloud, distinct from the one associated with the IRAS\,4A system.

Despite numerous studies on this system, the intricate morphology of the outflows has posed challenges in definitively attributing them to the respective protostars. Consequently, a comprehensive understanding of the broader scenario remains elusive. Achieving a more coherent picture demands advancements in several aspects, including higher spatial and spectral resolution, enhanced sensitivity, wider field of view coverage, and the use of diverse tracers. 
In this paper, we have examined the morphology and kinematics of the jets and outflows towards the IRAS\,4A system, via the study of the line emission of three standard shock tracers (see above): SiO, H$_2$CO and HDCO. We employ these three tracers because a posteriori and after a fast check of the maps for this source we found that they offer complementary insights into the observed dynamics, with H$_2$CO and HDCO playing a crucial role in identifying the cavities and enriching our understanding of the system's dynamics. 
We aimed to conduct a comprehensive characterisation, pinpointing and associating distinct features with their respective protostars, using observations from the Large Program FAUST (Fifty AU STudy of the chemistry in the disc/envelope system of solar-like protostars; 2018.1.01205.L, PI: S. Yamamoto; \citealt{Codella2021}). 
The present article is structured as follows. 
In Sect. \ref{sec:observations} we describe the observations.
In Sect. \ref{sec:results} we present the molecular line maps, together with the main results of the analysis. 
In Sect. \ref{sec:discussion} we discuss the structure of the outflows in the IRAS\,4A protobinary system, as derived by the FAUST observations, and compare it with the interpretations in previous literature.
In Sect. \ref{sec:conclusions} we summarise the conclusions of this work.


%% file: sections/obs.tex
\label{sec:observations}

\begin{table*}
    \centering
    \caption{Spectral parameters of the SiO, H$_{2}$CO, and HDCO lines observed towards IRAS\,4A.}
    \begin{tabular}{cccccccccc}
    
         \hline
       $\mathrm{Transition}$  & $\mathrm{\nu \,} ^{(a)} $ & $\mathrm{E_{up} \,} ^{(a)}$ & $\mathrm{A_{ij} \,} ^{(a)}$ &  $\mathrm{S \, \mu ^{2} \,} ^{(a)}$ & Bandwidth & $\mathrm{dV} $ & FWHM FoV & $\mathrm{Beam \, (PA)}$ & $\mathrm{Chan. \, rms} $  \\
         &  (MHz) & (K) & (10$^{-4}$\,s$^{-1}$) &(D$^{2}$) & (MHz) & (km s$\rm ^{-1}$) & $\arcsec$& $\arcsec$ $\times$ $\arcsec$ ($\rm ^{\circ}$) &
         (mJy beam$\rm ^{-1}$) \\ \hline
         SiO(5--4) & 217104.98 & 31.3 & 5.20 & 48.8 & 58.6 & 0.2 & 29& 0.48 $\times$ 0.34 (87.8) & 0.81 \\
          H$_{2}$CO(3$_{0,3}$--2$_{0,3}$) & 218222.19 & 21.0 & 2.82 & 16.3 & 58.6 & 0.2 & 28.9 & 0.46 $\times$ 0.32 (-6.6) & 0.72  \\
         HDCO(4$_{1,4}$--3$_{1,3}$) & 246924.60 & 37.6 & 3.98 & 20.4 & 1875 & 1.4 & 25.5 & 0.37 $\times$ 0.27 (-17.4) & 0.40 \\
         \hline 
    \end{tabular}
    
    \label{tab1}
     \noindent
$^{(a)}$ Frequencies and spectroscopic parameters have been extracted from \cite{Muller2013} for SiO, \cite{Bocquet1996, Muller2017} for H$_{2}$CO, and \cite{Bocquet1999} for HDCO.
\end{table*}

The IRAS\,4A system was observed at 1.2\,mm with ALMA during its Cycle 6 operations, between October 2018 and September 2019, as part of the Large Program FAUST. 
The observations were performed in Band 6 using the 12-m array (C43-3 and C43-6 configurations, including 46 and 50 antennas, respectively) and the 7-m Atacama Compact Array (ACA). The baselines for the 12-m array were between 15.1\,m and 3.6\,km, probing angular scales from 0$\farcs$03 ($\sim$9\,au) to 8$\farcs$2 ($\sim$2400\,au).

The observations were centred at R.A. (J2000)\,=\, 03$^{\rm{h}}29^{\rm{m}}10^{\rm{s}}$.539, Dec. (J2000)$ \,= 31^{\circ}13'30''.92$ and the systemic velocity was set to V$_{\rm{lsr}}$\,=\,7\,km\,s$^{-1}$. 
Several spectral windows (spw) were placed within the spectral range 216–234 GHz and 243–262 GHz. In this work, we focus on two narrow spws covering the following lines: SiO(5-4) and H$_{2}$CO(3$_{0,3}$--2$_{0,3}$), and one large spw covering HDCO(4$_{1,4}$--3$_{1,3}$). The spectroscopic parameters of each line are reported in Table \ref{tab1}. The SiO and H$_{2}$CO spws have a bandwidth of 58.6 MHz ($\sim$80.9\,km\,s$^{-1}$ for SiO and $\sim$80.5\,km\,s$^{-1}$ for H$_{2}$CO spw) and a channel width of 141 kHz ($\sim$0.2\,km\,s$^{-1}$ for SiO and H$_{2}$CO spw). The window covering HDCO has a bandwidth of 1875 MHz ($\sim$2294.4\,km\,s$^{-1}$) and a channel width of 1.13 MHz ($\sim$1.4\,km\,s$^{-1}$).


The quasar J0237+2848 was used for bandpass and flux calibration, while J0336+3218 and J0328+3139 were used for phase calibration. The absolute flux calibration uncertainty is estimated to be <15\%. The data calibration was performed using the standard ALMA calibration pipeline with the Common Astronomy Software Applications package 5.6.1-8 (CASA\footnote{\url{https://casa.nrao.edu/}}, \citealp{CasaTeam2022}). An additional calibration routine\footnote{\url{https://help.almascience.org/kb/articles/what-are-the-amplitude-calibration-issues-caused-by-alma-s} \\ \url{-normalization-strategy}; see also Moellenbrock et al. in preparation} has been used to correct for the $T_{sys}$ and for spectral data normalisation. The spectrum of IRAS\,4A is particularly rich and the identification of line-free continuum channels had to be done by hand, outside the ALMA pipeline. Line-free channels were then averaged per-spectral window and used to create a continuum model for each ALMA configuration, and both amplitude and phase self-calibration were employed to correct for remaining tropospheric phase fluctuations, position offsets, and amplitude calibration differences between the execution blocks. For each ALMA configuration, two rounds of phase-only selfcal were performed, followed by two rounds of amplitude and phase selfcal once the model was sufficiently complete. The resulting improvement in the continuum dynamic range was typically a factor of 5--7. The continuum model was then subtracted from the line channels to produce continuum-subtracted visibilities. A similar self-calibration technique was then used to align data across configurations. The visibility data from the 12-m arrays were then combined to produce the continuum-subtracted line cubes. The resulting continuum-subtracted cubes were cleaned using H\"{o}gbom algorithm \citep{Hogbom1974} and natural weighting in the IRAM-GILDAS software package\footnote{http://www.iram.fr/IRAMFR/GILDAS/}. The data analysis was performed using the same package. The resulting beam size and rms are summarised in Table \ref{tab1}. The FWHM Field of View (FoV) is 29$''$ for SiO, 28.9$''$ for H$_{2}$CO, and 25.5$''$ for HDCO. The maps shown in the paper are not corrected for the primary beam attenuation, as we are more interested in the structures' morphology (rather than the flux) that was not affected by the primary beam attenuation. By not applying correction for primary beam attenuation, we ensure that features extending beyond the field of view, are not inadvertently excluded in our analysis. This approach allows us to preserve the integrity of the observed structures, facilitating a comprehensive understanding of their spatial distribution and morphology.

%% file: sections/results.tex
\label{sec:results}


We have imaged the emission from SiO, H$_{2}$CO, and HDCO towards the IRAS\,4A system. 
The comprehensive understanding of the observed dynamics originates from the combined insights derived from the analysis of the three tracers. For a concise overview of our findings, readers are directed to Section \ref{subsec:overview}.

\begin{figure*}
    \centering
    \includegraphics[width=0.98\textwidth,  
    ]{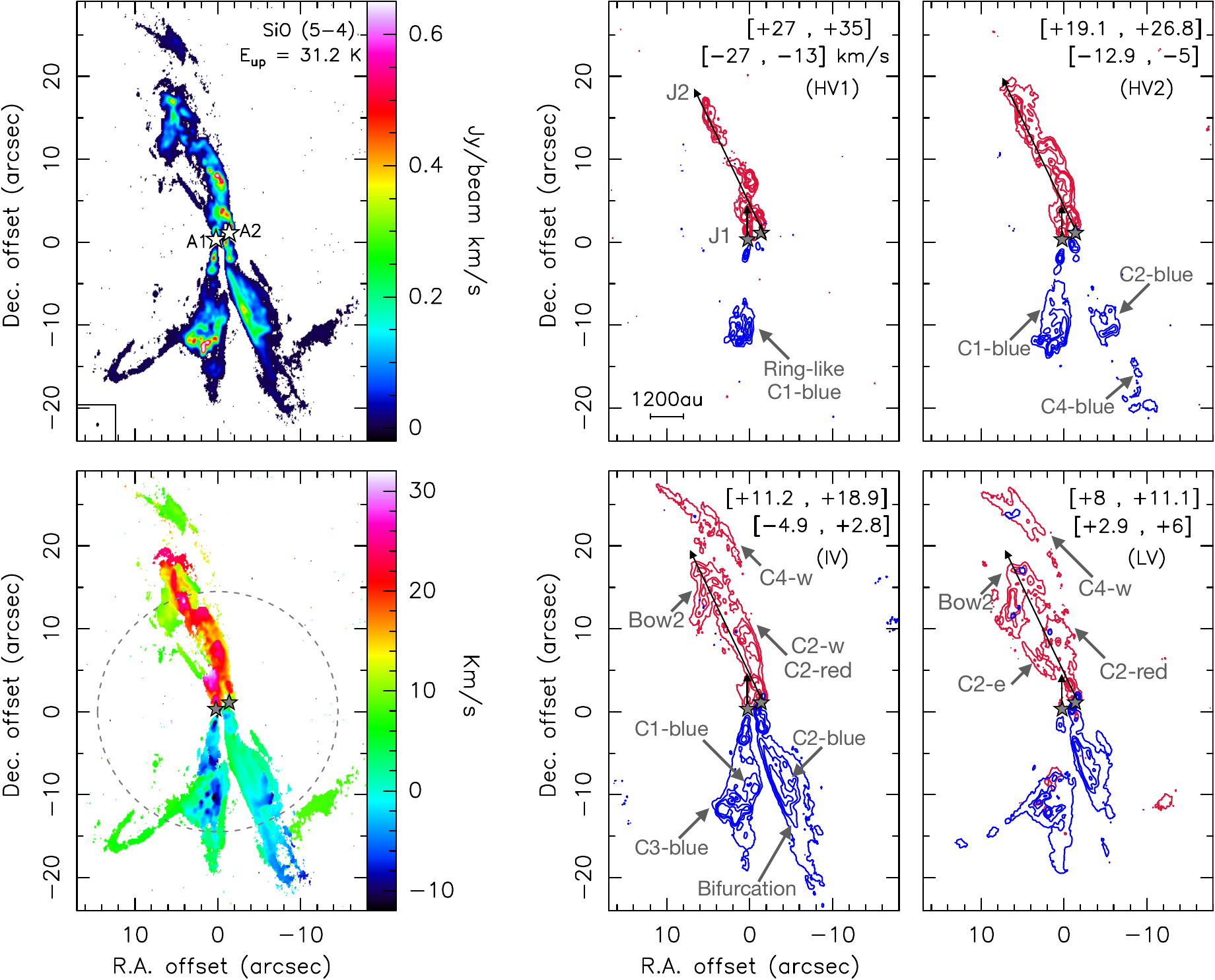} 
     
    \caption{Integrated intensity (moment\,0; top left) map, velocity map (moment\,1; bottom left) and channel maps (right panels) of SiO(5-4) towards the NGC\,1333 IRAS\,4A system. In the moment\,0 and moment\,1 panels the SiO emission is integrated between -27 and 35\,km\,s$^{-1}$ and a 5$\sigma$ threshold cut was applied \protect \footnotemark  ($\sigma$\,=\,0.81\,mJy\,beam$^{-1}$\,km\,s$^{-1}$). The channel maps correspond to four different velocity intervals: High-Velocity 1 (HV1), High-Velocity 2 (HV2), Intermediate-Velocity (IV) and Low-Velocity (LV). The intervals and their corresponding labels are shown at the upper right corners of each channel maps panel. The contour levels are [5, 25, 40, 60, 100]\,$\sigma$ for the redshifted HV1 regime and at [5, 25, 50, 100]\,$\sigma$ for the other regimes with $\sigma$\,= (1.3\,mJy\,beam$^{-1}$\,km\,s$^{-1}$, 1.5\,mJy\,beam$^{-1}$\,km\,s$^{-1}$, 1.7\,mJy\,beam$^{-1}$\,km\,s$^{-1}$, 0.9\,mJy\,beam$^{-1}$\,km\,s$^{-1}$, respectively). The positions of IRAS 4A1 and IRAS 4A2 are marked with stars. The jets from each source are marked with arrows. The synthesised beam is depicted in the lower-left corner of the top-left panel. The FWHM of the Field of View is depicted with a dashed grey circle in the lower-left panel.}
    
   \label{sio-mom-chan}
\end{figure*}


\begin{figure}
    \centering
  
    \includegraphics[width=0.48\textwidth]{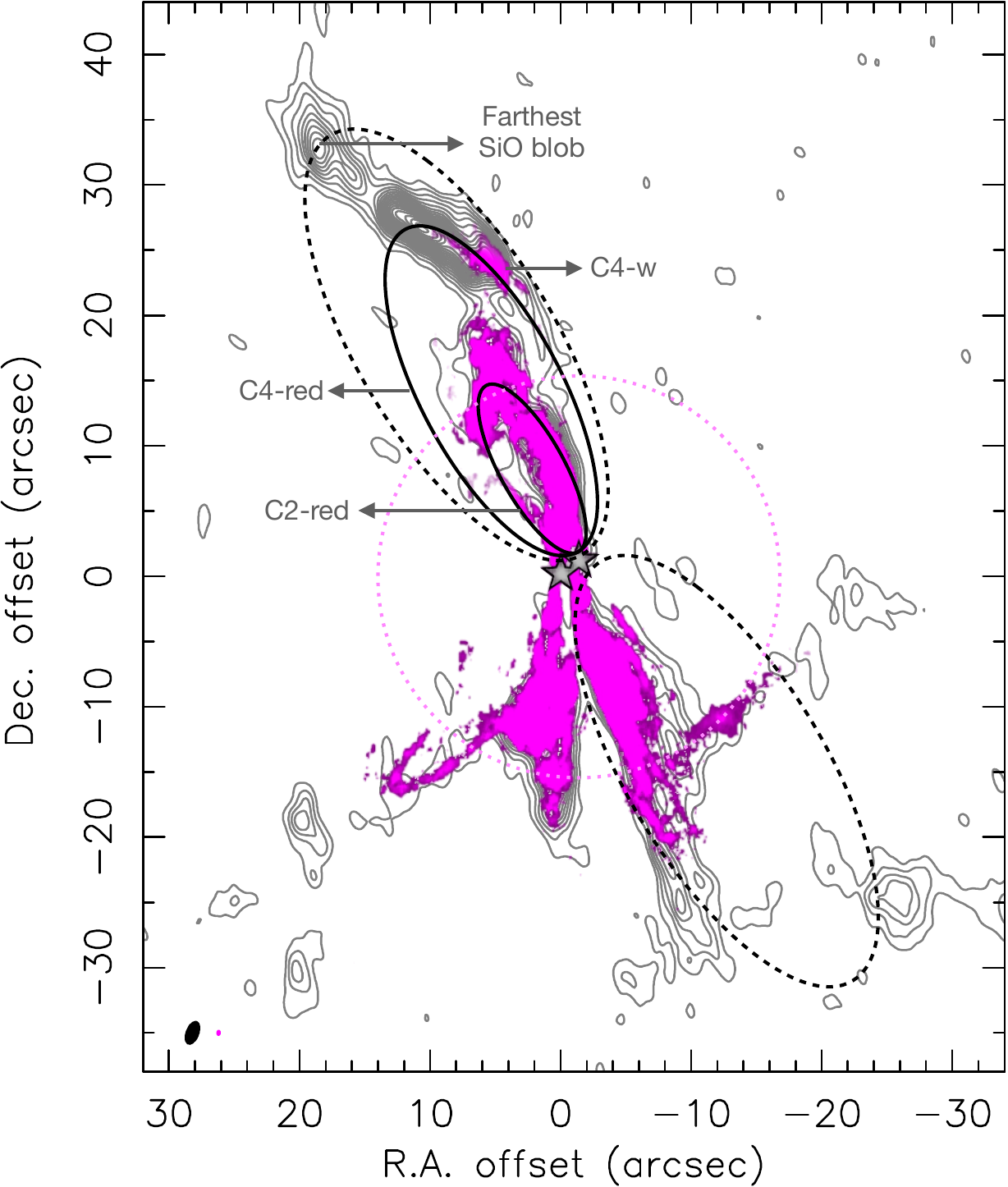} 
    
    \caption{SiO(2-1) integrated intensity contours from \protect \cite{DeSimone2022} with superimposed SiO(5-4) integrated intensity map from this work in colours. The outflow cavities emanating from IRAS\,4A2 and identified in this work are represented with solid ellipses. The possible third event ("Hypothetical cavity") emanating from IRAS\,4A2 is represented with a dashed ellipse. The FWHM of the Field of View in our observations is depicted with a dotted magenta circle. The synthesised beams are depicted in the lower-left corner.}
    
   \label{sio-choi}
\end{figure}

\subsection{SiO emission}

Figure \ref{sio-mom-chan} illustrates a thorough view of the SiO emission in six distinct panels, depicting the spatial and velocity features of the studied region. 
The upper left panel shows the SiO(5--4) integrated intensity map (moment\,0), the lower left panel shows the SiO velocity map (moment\,1) both integrated between -27 and 35\,km\,s$^{-1}$, and the remaining four panels show the integrated channel maps, each representing SiO emission at distinct velocity intervals. 
The SiO moment 0 map reveals different features, including bow-shocks, jets, and outflow, alongside the finger-like structures previously reported by \cite{DeSimone2022} in the south.
The SiO emission in the figure spans from -27\,km\,s$^{-1}$ ($\delta$V= -34\,km\,s$^{-1}$; V$_{\rm {lsr}}$=\,7\,km\,s$^{-1}$) to 35\,km\,s$^{-1}$ ($\delta$V= +28\,km\,s$^{-1}$). The narrow spectral window limits detection beyond 35\,km\,s$^{-1}$, however, the emission is expected to be relatively low beyond this velocity threshold. The moment 1 map indicates that features located north of the protostars are characterised by redshifted velocities, while those situated south exhibit blueshifted velocities. The integrated channel maps in the figure correspond to four different velocity intervals, namely the high-velocity 1 (HV1; $\delta$V= -34\,(+28)\,--\,$\pm$20\,km\,s$^{-1}$), high-velocity 2 (HV2; $\delta$V= $\pm$12\,--\,$\pm$20\,km\,s$^{-1}$), intermediate-velocity (IV; $\delta$V= $\pm$4\,--\,$\pm$12\,km\,s$^{-1}$), and low-velocity intervals (LV; $\delta$V= $\pm$1\,--\,$\pm$4\,km\,s$^{-1}$).

\subsubsection{High Velocity (HV1 and HV2)}

\textbf{HV1}: In the redshifted HV1 regime, a north-south jet (J1) stemming from source 4A1 is evident. The jet shows bullet-like structures and reaches almost $\sim$5$''$ ($\sim$1500\,au) from the protostar. At distances of 12$''$--18$''$ ($\sim$3500--5300\,au) from 4A2, another jet (J2) is visible, clearly directed towards it, and distinguished by the presence of identifiable bullets along this feature. A zoom-in onto the innermost region showing of J1 from J2 is depicted in Fig. \ref{sio-jets}.

On the other hand, in the blueshifted emission, we detect a ring-like feature, directed almost N-S, associated with the first blueshifted outflow cavity of 4A1 (C1-blue).
\newline
\textbf{HV2}: In the panel corresponding to the HV2 interval, the emission starts to overlap at the north. Conversely, in the south, the components attributed to 4A1 can be distinguished from those of 4A2. At $\sim$13$''$ ($\sim$3800\,au) to the south of 4A2, we identify a prominent bow-shock structure associated with an outflow cavity of 4A2 (C2-blue). Additionally, there is observable emission around $\sim$23$''$ ($\sim$6700\,au) to the south of 4A2, which is likely associated with another cavity (C4-blue; See Sect. \ref{subsec:h2co}). 



\begin{figure}
    \centering
    \includegraphics[width=0.48\textwidth]{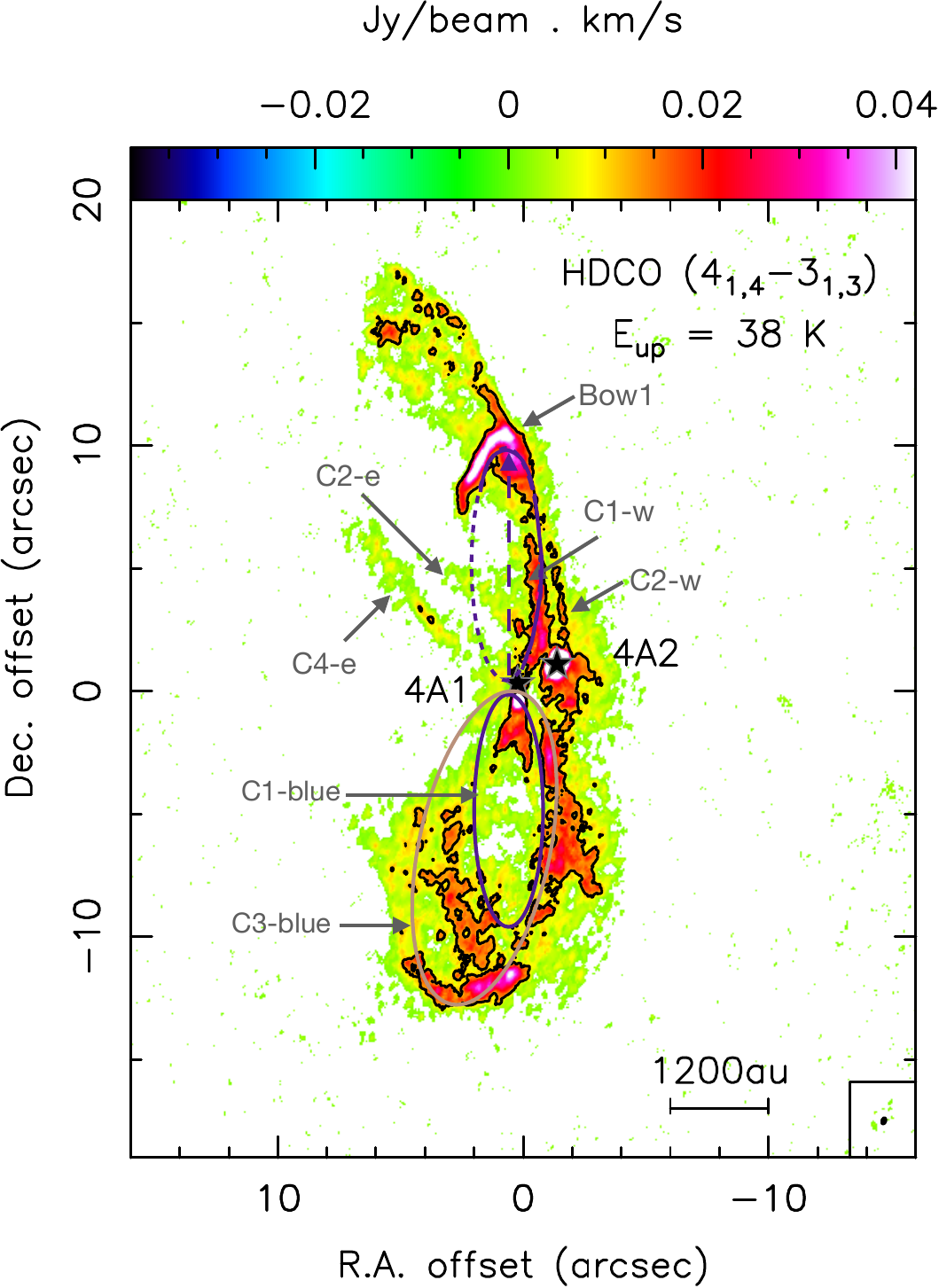} 

    \caption{Moment\,0 map of HDCO integrated between -3.6 and 20.1\,km\,s$^{-1}$. A 3$\sigma$ threshold cut was applied ($\sigma$\,=\,0.40\,mJy\,beam$^{-1}$\,km\,s$^{-1}$). The contour levels are at 30\,$\sigma$ with $\sigma$\,= 0.40\,mJy\,beam$^{-1}$\,km\,s$^{-1}$. The positions of IRAS\,4A1 and IRAS\,4A2 are marked with stars. The features mentioned in the text are labelled in grey. The solid ellipses represent the IRAS\,4A1 outflow cavities observed in HDCO. The synthesised beam is depicted in the lower-right corner.}
    
   \label{hdco-mom}
\end{figure}

\subsubsection{Intermediate Velocity (IV)}

Transitioning to the redshifted intermediate velocity (IV) regime, shown in the bottom left panel of Fig. \ref{sio-mom-chan}, the SiO emission presents a complex overlap of the structures originating from 4A1 and 4A2, making interpretation difficult. These features are better discerned in HDCO and will be discussed further (See Sect. \ref{subsec:hdco}). However, with the detection of a bow-shock at $\sim$18$''$ ($\sim$5300\,au) from 4A2 (Bow2), which is also perpendicular to the trajectory of its jet 4A2 (J2), we observe that the extended northeast-southwest outflow cavity wall (C2-w) reaches Bow2. This observation suggests that we are tracing the high-velocity component of the 4A2 outflow (C2-red; further explanation is provided in the subsequent sections). The outflow shows an S-shape bending within 4$''$ ($\sim$1200\,au) from IRAS\,4A2 that was also observed in previous studies \citep{Santangelo2015, Chuang2021}. It is worth noting that Bow2 was present in the CS maps of \cite{Taquet2020}, but was not explicitly identified by the authors. Additionally, an arc-like feature appears at approximately 20--25$''$ ($\sim$5900--7300\,au) from 4A2 (C4-w). 
This feature is coincident with what was interpreted as a sharp bend (also referred to as deflection or directional variability) by \cite{Choi2005}, which has also been observed by \cite{yildiz2012, DeSimone2022}, as illustrated in Fig. \ref{sio-choi}. 


Initially, this structure was attributed to a deflection of the 4A2 outflow caused by a dense core \citep{Choi2005}. 
However, we suggest that it is rather linked with the LV component of the 4A2 outflow (C4-red; see Sect. \ref{subsec:h2co}), being the remnant of its western cavity wall. The emission of this outflow is better discerned with H$_{2}$CO. 
Within the same velocity interval, the blue-shifted emission shows the presence of two bow-shocks directed towards 4A1, one in the N-S direction reaching 10$''$ ($\sim$3000\,au) from the protostar and the other with a PA of $\sim$-12$^{\circ}$ reaching $\sim$13$''$ ($\sim$3800) from it. Although less clear in SiO, the bow-shocks are better observed in the other tracers, and they illustrate two outflow cavities driven by 4A1 (C1-blue and C3-blue). To the south of 4A2, distinguishing the emission becomes more intricate, and we likely observe two outflows extending $\sim$13$''$ (C2-blue) and $\sim$23$''$ (C4-blue) from the protostar. The 4A2 outflow resembles that of HH\,46/47 \citep{Arce2013} showing a bifurcation along its eastern wall.

\subsubsection{Low Velocity (LV)}

In the redshifted low-velocity regime, the emission is dominated by the components linked with 4A2. Indeed,  prominent outflow cavity walls directed towards 4A2 are distinctly visible towards the west and the east (i.e. C2-w, C2-e and C4-w) together with the Bow2 bow-shock. We suggest that C2-w, C2-e and Bow2 delineate the HV component of the 4A2 outflow (C2-red) and that C4-w is linked with the LV component of the 4A2 outflow cavity wall (C4-red; See Sect. \ref{subsec:h2co} for further explanation)

Lastly, it is noteworthy that in the moment\,0 map of SiO, a bow-like feature is discernible to the east of the 4A1 blueshifted outflow (C1-blue). This structure was previously identified by \cite{DeSimone2022} at lower angular resolution, where it resembled a finger-like structure similar to the one observed at the west of the A2 outflow. The authors proposed these fingers to be formed by a train of shocks due to an expanding bubble crashing against NGC\,1333. The higher resolution of the ALMA data reveals that the finger-like structure at the west of the 4A1 outflow has an arc-like shape at its end. However, since there is no apparent association with any specific outflow emanating from 4A1 or 4A2, we postpone a dedicated investigation of these fingers to future work.


\begin{figure*}
    \centering
    \includegraphics[width=0.48\textwidth]{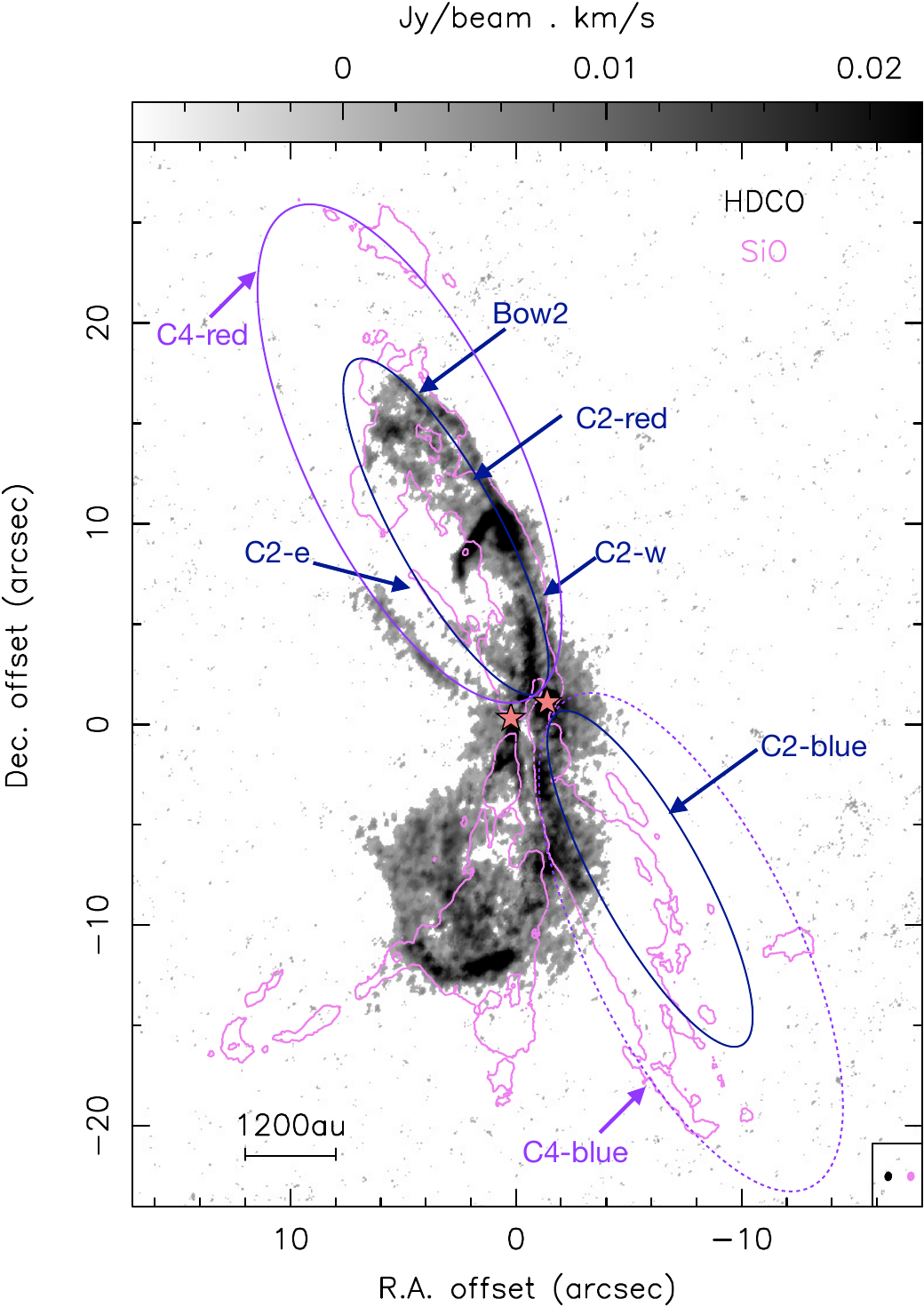} 
      \includegraphics[width=0.48\textwidth]{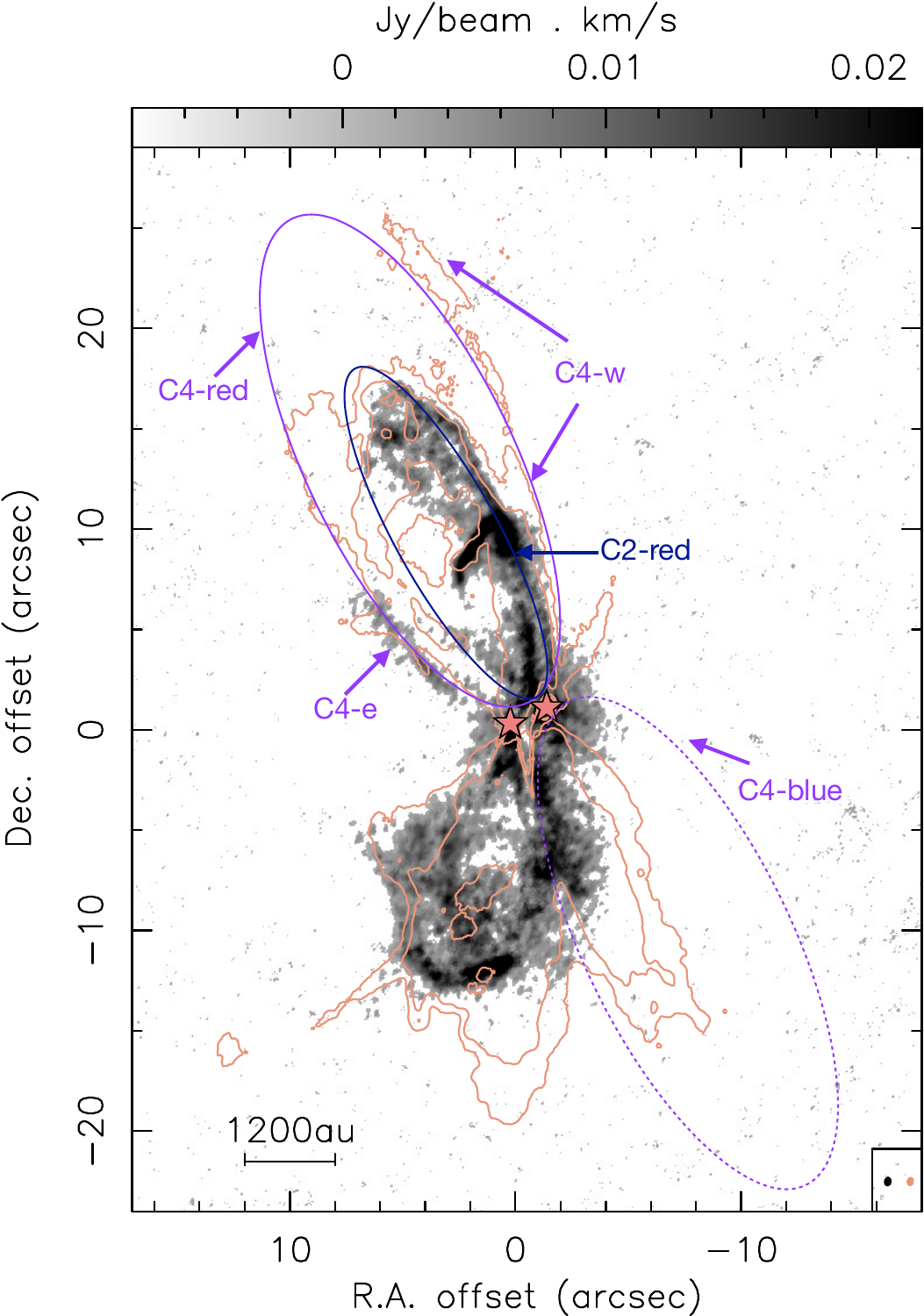}
    \caption{Left: Moment\,0 map of HDCO in colour scale with superimposed SiO contours at 5$\sigma$ ($\sigma$\,=\,0.81\,mJy\,beam$^{-1}$\,km\,s$^{-1}$). The positions of IRAS 4A1 and IRAS 4A2 are marked with stars. The HV component of the 4A2 outflow is represented by a navy blue ellipse. The synthesised beams are depicted in the lower-right corner. Right: Moment\,0 map of HDCO in colour scale with superimposed H$_{2}$CO 5$\sigma$ and 30$\sigma$ redshifted LV contours in coral ($\sigma$\,=\,1.5\,mJy\,beam$^{-1}$\,km\,s$^{-1}$). The positions of IRAS\,4A1 and IRAS\,4A2 are marked with stars. The LV component of the 4A2 outflow is represented by a purple ellipse. The synthesised beams are depicted in the lower-right corner.}
    
   \label{hdco-sio}
\end{figure*}

\begin{figure*}
    \centering
    \includegraphics[width=0.98\textwidth]{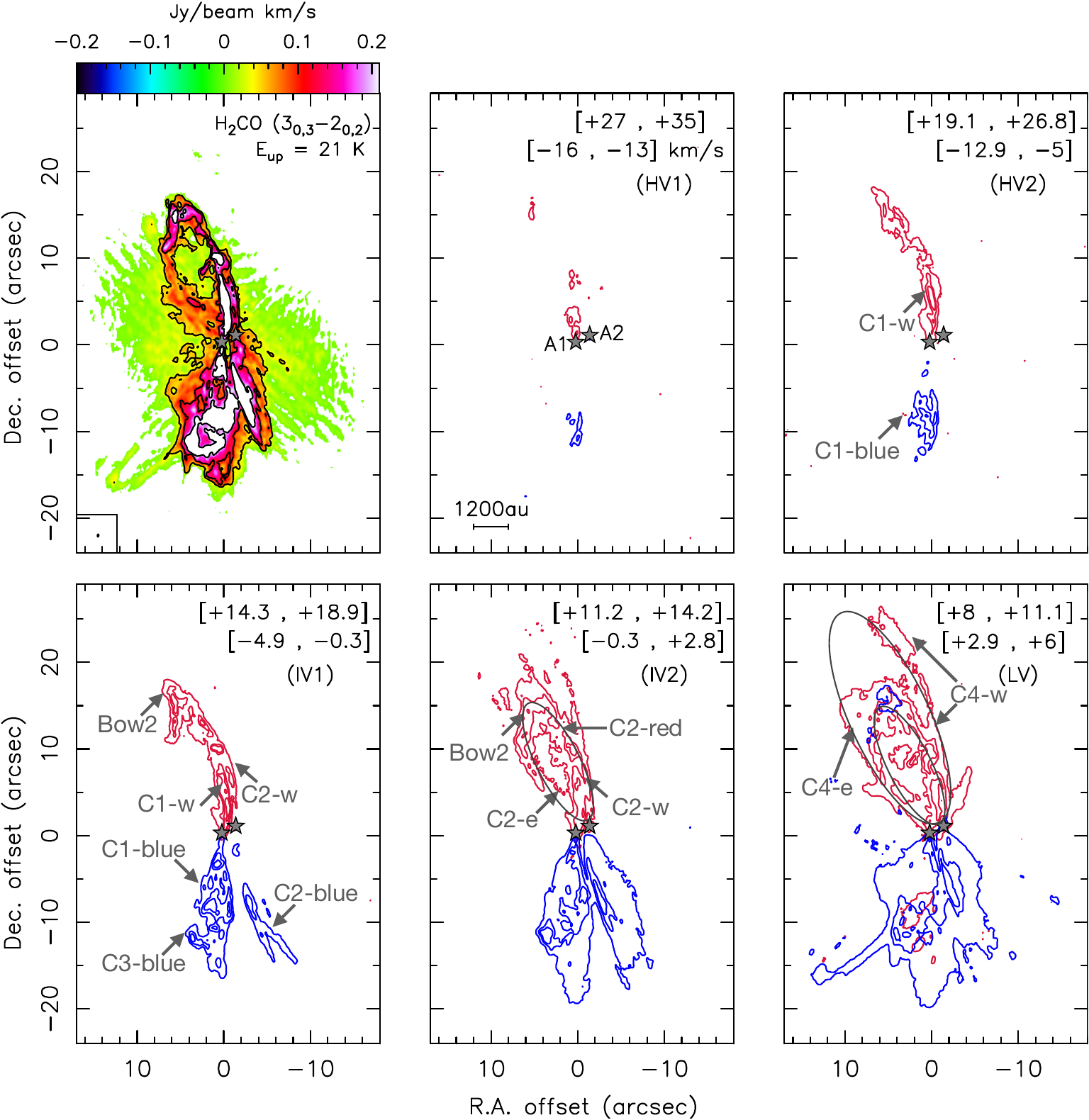} 

        \caption{Integrated intensity (moment\,0; top left) map and channel maps (right panels) of H$_{2}$CO towards the NGC\,1333 IRAS\,4A system. In the moment\,0 the H$_{2}$CO emission is integrated between -16 and 35\,km\,s$^{-1}$ and a 15$\sigma$ threshold cut was applied ($\sigma$\,=\,0.72\,mJy\,beam$^{-1}$\,km\,s$^{-1}$). The channel maps correspond to five different velocity intervals High-Velocity 1 (HV1), High-Velocity 2 (HV2), Intermediate-Velocity (IV1 and IV2) and Low-Velocity (LV). The intervals and their corresponding labels are shown at the upper right corners of each channel map panel. The contour levels are at [5, 25]\,$\sigma$ for HV1, [15, 50, 100]\,$\sigma$ for the blueshifted emission in the HV2 and IV1 velocity intervals, [5, 35, 50, 100]\,$\sigma$ for the redshifted emission in IV1, and [5, 50, 100]\,$\sigma$ for the others, with $\sigma$\,= 0.9\,mJy\,beam$^{-1}$\,km\,s$^{-1}$, 1.1\,mJy\,beam$^{-1}$\,km\,s$^{-1}$, 1.2\,mJy\,beam$^{-1}$\,km\,s$^{-1}$, 1.0\,mJy\,beam$^{-1}$\,km\,s$^{-1}$, 1.5\,mJy\,beam$^{-1}$\,km\,s$^{-1}$, respectively). The positions of IRAS\,4A1 and IRAS\,4A2 are marked with stars. The solid ellipses in the last two panels represent the C2-red and C4-red cavities. The synthesised beam is depicted in the lower-left corner of the top-left panel.}

   \label{h2co-mom-chan}
\end{figure*}


\subsection{HDCO emission}
\label{subsec:hdco}

In Fig. \ref{hdco-mom} we present the moment\,0 map of HDCO integrated between (-3.6 and 20.1\,km\,s$^{-1}$). 
HDCO emission delineates the outflow cavities towards the IRAS\,4A system, helping to better discern different components, particularly to the north of the protostar, despite some overlap in emission. Notably, HDCO emission reveals, for the first time, a bow-like feature at $\sim$10$''$ from 4A1 (Bow1) perpendicular to its N-S jet. Stronger emission contours also show two curved features (C1-w and C2-w), one aligned with 4A1 and another with 4A2. For a detailed inspection, we show the channel map of the HDCO redshifted emission towards the north (V\,=\,8.4--17.8\,km\,s$^{-1}$; Fig. \ref{hdco-chan}). We note that C1-w is emanating from A1 and has a higher velocity compared with the other features (including C2-w). This feature (C1-w) extends up to $\sim$10$''$ from 4A1 reaching the bowshock Bow1, suggesting that Bow1 and C1-w delineate the 4A1 cavity, oriented in the N-S direction (C1-red). Conversely, C2-w emission extends beyond 10$''$ (see also Fig. \ref{hdco-chan}), forming a V-shape emerging from 4A2, together with C2-e (detected in SiO). The overlap between the SiO and HDCO emission (Fig. \ref{hdco-sio}), shows that these two features and the Bow2 bow-shock observed in SiO are actually tracing the same outflow cavity, which we named C2-red, where C2-w is the western cavity wall and C2-e is the eastern. Further, emission beyond 10$''$ from the protostars aligns with the western wall of C2-red, suggesting an association with A2. Notably, while C2-w and C2-e belong to the same cavity, C2-w exhibits a higher velocity by almost 3\,km\,s$^{-1}$, indicating possible cavity rotation. Moving southward, an extended cavity from 4A1 with PA$\sim$-12$^{\circ}$ east (C3-blue) and part of the eastern wall of the cavity from 4A2 can be observed. Intriguingly, within the 4A1 cavity, another “nested” cavity, C1-blue, directed almost N-S, is noticeable (see Sect. \ref{subsec:h2co} for further discussion).  It is worth noting that while C1-blue has C1-red as a redshifted counterpart, C3-blue lacks a counterpart, suggesting it could be exiting the cloud or hidden by the emission of the 4A2 cavity.

\subsection{H$_{2}$CO emission}
\label{subsec:h2co}

Figure \ref{h2co-mom-chan} presents the maps of the IRAS\,4A system using H$_{2}$CO as a molecular tracer. The H$_{2}$CO emission is illustrated in six panels, the upper left panel shows the H$_{2}$CO moment\,0 map, and the remaining five panels show the integrated channel maps. H$_{2}$CO effectively maps the distribution of the bulk gas within the IRAS\,4A system; it also traces the intricate outflow cavities originating from both the 4A1 and 4A2 sources, as well as the jet from 4A1. The emission in the figure spans from -16\,km\,s$^{-1}$ ($\delta$V= -23\,km\,s$^{-1}$; V$_{\rm {lsr}}$=\,7\,km\,s$^{-1}$) to 35\,km\,s$^{-1}$ ($\delta$V= +28\,km\,s$^{-1}$). The channel maps velocity intervals correspond to five different velocity intervals, the HV2 and IV are as in Fig. \ref{sio-mom-chan}, while for the HV1 regime ($\delta$V= -23\,(+28)\,--\,$\pm$20\,km\,s$^{-1}$). To better illustrate the features of the IV regime, we split it into IV1 and IV2. We note that the SiO map encompasses a slightly broader range of velocities compared to H$_2$CO, which is indicative of the latter exhibiting dimmer emissions that remain undetected at the highest velocities.

\subsubsection{High Velocity (HV1 and HV2)}

\textbf{HV1:} In the HV1 panel, the H$_{2}$CO emission displays the presence of the previously identified north-south jet originating from 4A1 (J1). 

\noindent\textbf{HV2:} In the redshifted HV2 interval, we detected the previously observed curved feature C1-w from 4A1 (i.e. the western wall of the redshifted cavity), consistent with our HDCO observations. Additionally, we identified another feature oriented northeast-southwest, pointing towards 4A2 and aligned with the western wall of C2-red (seen in SiO). Transitioning to the blueshifted regime, we identify a distinct N-S outflow cavity from 4A1 (C1-blue). 

\subsubsection{Intermediate Velocities (IV1, IV2)}

\begin{figure*}
    \centering
    \includegraphics[width=0.98\textwidth,  
    ]{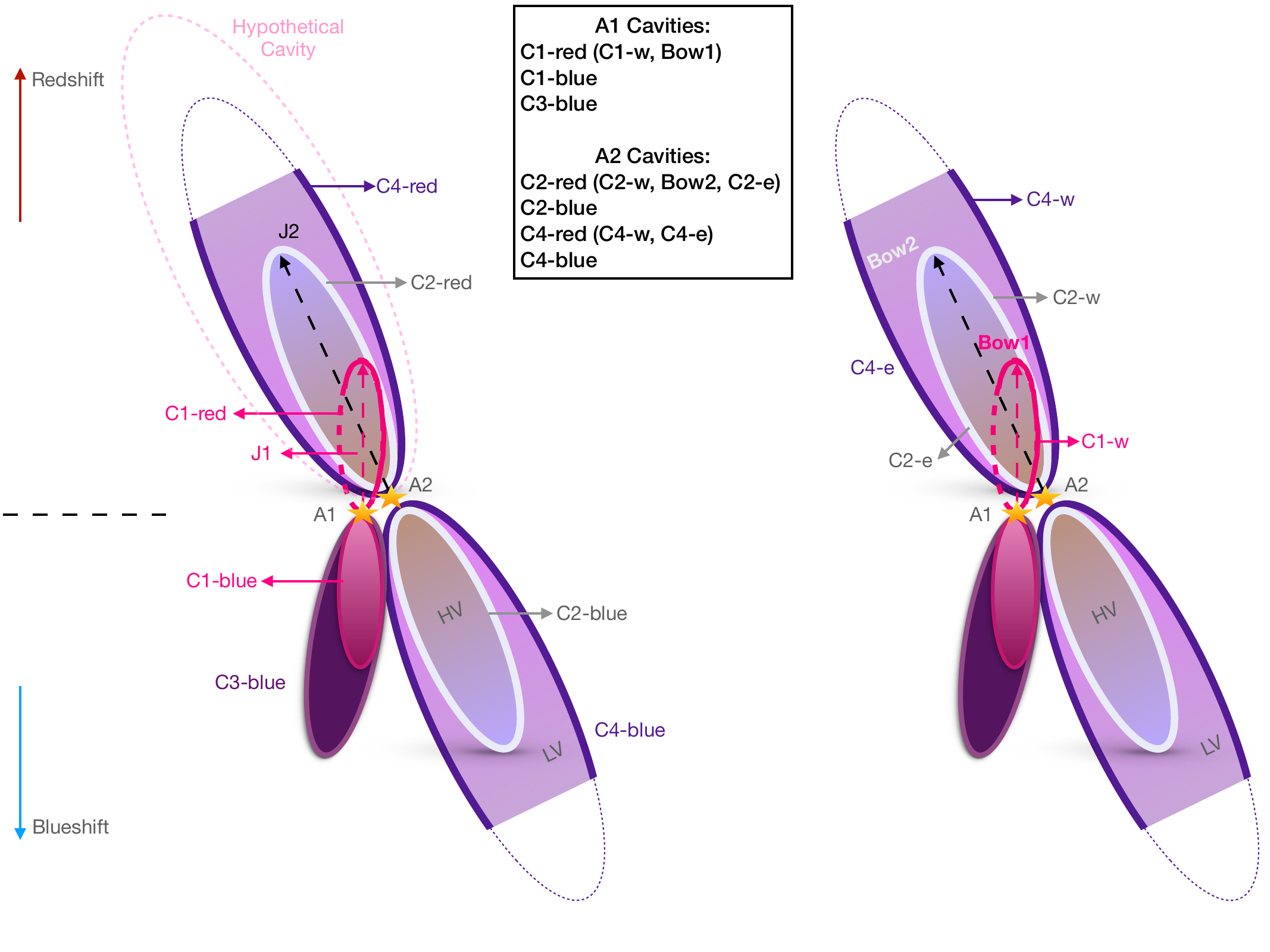}     
    \caption{Sketch illustrating the comprehensive perspective of the jets and outflows towards the IRAS\,4A system as proposed in this study. The left diagram incorporates the annotations of the jets and various cavities proposed in this work. The right diagram highlights the annotations of the distinct features of the redshifted cavities, including the cavity walls (east "e" and west "w") and the bow shocks ("Bow"), as discussed and referenced in the text.}
    
   \label{fig:sketch}
\end{figure*}

Moving further into the blueshifted IV1 regime, as in SiO, clear differentiation emerges between the distinct outflow cavities, attributed to 4A1 and 4A2. 
Within the 4A1 cavity across both IV1 and IV2 regimes, akin to HDCO observations, we observe two bow-like features. One is nearly perpendicular to the N-S jet direction (C1-blue), while the other is tilted about 12$^{\circ}$ eastward (C3-blue), affirming the presence of dual cavities from this source (see also the sketch in Fig. \ref{fig:sketch}). Additionally, the cavities manifest at different velocities; C1-blue is observable starting from -12.5 km/s, whereas the bowshock of C3-blue is discernible at -6.9 km/s, with the cavity becoming more apparent at lower velocities (See Fig \ref{h2co-chan-blue}). To the south of 4A2, at least one outflow is observed.

The interpretation of the emission in the redshifted regime presents considerable complexity, as in the other tracers. The 4A1 outflow cavity (C1-red) is detected. The western lobe (C1-w) extends up to 10$''$ from 4A1, while we see only 3$''$ from the eastern lobe. The features delineating C2-red (i.e. C2-w, Bow2 and C2-e) are also detected. As in HDCO, the western cavity wall C2-w is observed at higher velocities, being 7\,km\,s$^{-1}$ faster than eastern C2-e (as shown in the detailed channel map \ref{h2co-chan-red}) suggesting a possible cavity rotation. 

\subsubsection{Low Velocity (LV)}

The redshifted LV regime is primarily characterised by the outflows originating from 4A2, notably the extended northeast-southwest cavity (C2-red) detected in SiO. Alongside this cavity, denoted as C2-red, we identify another structure labelled "C4-w," also observed in SiO, situated to the west of 4A2. This feature, stemming from IRAS\,4A2, extends approximately $\sim$25$''$ from the protostar, coinciding with the distinct directional variability previously documented by \cite{Choi2005, yildiz2012, DeSimone2022}. With H$_{2}$CO, C4-w is detected between $\sim$8 and 12\,km\,s$^{-1}$ (see the detailed channel map in Fig. \ref{h2co-chan-red}). Within this velocity range, we observe emergence of emission towards the northeast of Bow 2 and the east of C2-e (Fig. \ref{h2co-chan-red}). The overlap of the H$_2$CO LV contours with the HDCO moment 0 shows that the emission to the east of C2-e aligns with the HDCO feature C4-e, extending approximately $\sim$15$''$ from 4A2 (see Fig. \ref{hdco-sio}), suggesting that we are tracing the same feature. We propose that C4-w and C4-e together form an elliptical shape with a PA of $\sim$26$^{\circ}$, directed towards 4A2, implying they trace a second outflow system driven by 4A2 (C4-red), where C4-w represents the western wall and C4-e is the eastern. Hence, the structure previously thought to be a sharp bend (directional variability; \citealt{Choi2005, yildiz2012, DeSimone2022}) is actually part of the western cavity wall of C4-red. The cavity is not entirely detected, likely due to filtering. In Fig. \ref{hdco-sio}, we depict the “potential” blue counterpart of C4-red with a purple dashed ellipse. This ellipse aligns with the SiO emission observed at 23$''$ from 4A2, indicating that it may represent a portion of the counterpart of C4-red. However, it is challenging to differentiate it clearly from C2-blue, and its western wall is showing very weak emission (see moment 0 map of H$_2$CO).


\renewcommand{\arraystretch}{1.8}

\begin{table*}
    \begin{center}
        \caption{Summary of the components of the outflows emanating from IRAS\,4A1 and IRAS\,4A2  and detected in the SiO, H$_2$CO and HDCO maps presented in this work. The systemic velocity of the system is V$_{sys}$\,=\,7\,km\,s$^{-1}$ \citep{Su2019}.}
\resizebox{\textwidth}{!}{
{\begin{tabular}{ccccccccccc}
    \hline \hline  
    Source & Component & \multicolumn{3}{c}{Molecules} & Extension & \multicolumn{3}{c}{Velocity Interval (km\,s$^{-1}$)} & Notes & \makecell{Present in \\ the maps of} \\  
 &  & SiO & H$_{2}$CO & HDCO &  $''$ & SiO & H$_{2}$CO & HDCO &  &  \\  \hline

\multirow{5}{2em}{4A1}  & J1 & \checkmark & \checkmark & x  &  5 & [+25 , +35] & [+25 , +35] & x & N-S Jet of 4A1 & a, c, d, f, g \\

    & Bow1 & x & \checkmark & \checkmark & 10* & x & [+8 , 13] & [+8 , +13] & Bow-shock created by J1& a \\

    & C1-red & \checkmark & \checkmark & \checkmark  & 10 & [+10 , +22.5] & [+8.3 , +22.5] &[+8.3 , +19] & \makecell{N-S redshifted outflow cavity of 4A1 \\ includes Bow1 and C1-w }& a, e, g\\

    & C1-blue & \checkmark & \checkmark & \checkmark  & 10 & [-24 , +4.7] & [-14 , +6] & [-1.2 , +5.8] & N-S blueshifted outflow cavity of 4A1 & a, c, e, f, g\\

    & C3-blue & \checkmark & \checkmark & \checkmark  & 13 &  [-6.5 , +5.5] & [-5 , +6] & [+1 , +7] & \nth{2} blueshifted outflow cavity of 4A1, PA $-$12$^{\circ}$ & a, d, e\\

     \hline \hline
 \multirow{6}{2em}{4A2}  & J2 & \checkmark & x & x & 18* & [+23 , +35] & x & x & Component connected with the NE-SW Jet of 4A2 & a \\

         & Bow2 &\checkmark & \checkmark & x  &  18* & [+7.2 , +19] & [+8.3 , +19] & x& Bow-shock created by J2 & a, f \\

         & C2-red  & \checkmark & \checkmark & \checkmark & 18 & [+8 , +21] & [+8.3 , +19] &[+8 , +13] & \makecell{\nth{1} redshifted outflow cavity of 4A2, PA 29$^{\circ}$ \\ includes Bow2, C2-w and C2-e} & a, b, c, d, e, g\\

         & C2-blue & \checkmark & \checkmark & \checkmark& 13 & [-11 , +6] & [-4 , +6] & [+1 , +7]& \nth{1} blueshifted outflow cavity of 4A2, PA 29$^{\circ}$  & a, g\\

          & C4-red & \checkmark & \checkmark & x & 25 & [+7.2 , +12.8] & [+8.3 , +12.5] & x& \makecell{\nth{2} redshfited outflow cavity of 4A2, PA 26$^{\circ}$ \\ includes C4-w and C4-e} & a\\

         & C4-blue &\checkmark & weak & x & 23 & [-5.5 , //] & x & x& \nth{2} blueshifted outflow cavity of 4A2, PA 26$^{\circ}$ & a \\

         \hline
         \hline
    \end{tabular}}}
\noindent
\textit{\textbf{References: }} (a) This work, (b) \cite{Choi2005}, (c) \cite{Santangelo2015}, (d) \cite{Ching2016}, (e) \cite{Su2019}, (f) \cite{Taquet2020}, \newline (g) \cite{Chuang2021}. \newline \noindent * These values represent the location of the feature from the corresponding protostar, rather than the extension.
\label{tab:summary-components}
    \end{center}

\end{table*}

\subsection{Overview of the outflows and associated molecules}
\label{subsec:overview}

To simplify and present a concise overview, we have summarised the results in the sketch of Fig. \ref{fig:sketch} and listed the found components and their major properties in Table \ref{tab:summary-components}. A comprehensive figure is also shown in the appendix (Fig. \ref{all-mom-ellipses}, wherein all the features identified in this work are overlaid on the molecular integrated velocity maps of SiO, H$_2$CO and HDCO, with corresponding labels.

Overall, our observations reveal the presence of two outflows emanating from 4A1 and another two outflows emanating from 4A2.
As summarised in the introduction, this is the first time that four outflows are identified in the region: each protostar, 4A1 and 4A2, is responsible for two of them.
This conclusion is built on the various components detected in our maps in different lines at different velocities, as briefly listed here to guide the reader through their detailed description in the following sections.

\smallskip
\noindent
\textit{Jets J1 and J2:}
The high-velocity maps of SiO and H$_2$CO reveal the presence of two red-shifted jets, J1 and J2, north of 4A1 and 4A2, emanating from 4A1 and 4A2, respectively.

\smallskip
\noindent
\textit{Bow-shocks Bow1 and Bow2:}
The lower velocity maps of H$_2$CO and HDCO show a red-shifted bow-shock, Bow1, associated with J1, while those of SiO and  H$_2$CO show another bow-shock, Bow2, associated with J2. 

\smallskip
\noindent
\textit{A1 Cavities C1-red; C1-blue; C3-blue:}
Despite the clear detection of only one jet, J1, emanating from 4A1, our SiO, H$_2$CO and HDCO maps show the presence of one red-shifted cavity in the north, C1-red, and two blue-shifted cavities in the south, C1-blue and C3-blue.
C1-red and C1-blue are aligned almost North-South, indicating that they are part of the same bipolar outflow system.
Likewise, the alignment of the jet J1, the bow-shock Bow1 and the cavity C1-red suggest that they all belong to the same ejection event that created them, which reaches a distance from 4A1 of about 10$''$.

A second cavity, labelled C3-blue, extending to the south of 4A1, is distinctly visible, particularly in the HDCO emission. This cavity exhibits an eastward tilt in comparison to C1-blue. Notably, it lacks a corresponding red-shifted counterpart to the north. We suggest that the absence of a redshifted counterpart might be influenced by the overlap with the redshifted cavity of 4A2. This overlap could potentially obscure the cavity associated with 4A1, or even interact with it, dispersing its emission. As for the prospects of detection, it is plausible that the redshifted counterpart would be observable at larger distances, beyond the influence of the 4A2 cavity's obscuration, however, more observations would be required to investigate this point.


\smallskip
\noindent
\textit{A2 Cavities C2-red, C2-blue, C4-red and C4-blue:}
We identify two red-shifted cavities emanating from 4A2 in the north, C2-red, C4-red, and we associate them with two blue-shifted cavities in the south, C2-blue and C4-blue, although the separation is less clear.
The C2-red is visible in the intermediate-velocity maps of SiO, H$_2$CO and HDCO, while the C4-red appears in the low-velocity maps of SiO and H$_2$CO.
Both cavities, C2-red and C4-red, have an aligned blue-shifted counterpart in the south, C2-blue and C4-blue, respectively.
The C2-red is aligned with the jet emanating from 4A2, J2, with a position angle (PA) of $\sim$29$^{\circ}$. While C4-red is slightly tilted with a PA of $\sim$26$^{\circ}$. Finally, we notice that, in our interpretation, the emission observed in the SiO north of 4A2, previously interpreted as part of a deflected jet from 4A2 by \citep{Choi2005}, is, in fact, part of the cavity C4-red.

\smallskip
\noindent
\textit{Overall comments:}
The structures mentioned earlier are the outcome of assembling evidence that occasionally manifests in distinct tracers and at different velocities. Notably, components of the same cavity may not manifest simultaneously, particularly the eastern segments.

%% file: sections/discussion.tex
\label{sec:discussion}

\subsection{The different ejection events in the IRAS\,4A system}\label{subsec:disc-structure}

\smallskip
\noindent
\textit{4A1 outflow systems:}

Our observations disclose the existence of two discernible outflow systems (C1-red \& C1-blue, and C3-blue) originating from 4A1. 
The first (C1-red \& C1-blue) is directed N-S, spanning a distance up to 10$''$ both to the north and south of the protostar. 
However, the detection of the red-shifted eastern wall is somewhat limited, with only 3$''$ from it being observed. 
The cavity walls observed in this system are delineated by SiO, H$_2$CO, and HDCO, exhibiting a velocity of approximately $\sim$18\,km\,s$^{-1}$ in red. 

The second outflow system (C3-blue), driven by 4A1, is more extended, reaching nearly 15$''$ to the south of 4A1. Intriguingly, the counterpart of the 4A1 southeast cavity is absent in the northern region, implying its possible escape from the cloud, as in the case of HH\,46/47 \citep{Arce2013}. The blue-shifted lobe moves at a velocity of $\sim$15\,km\,s$^{-1}$. This outflow system is tilted by $\sim$12$^{\circ}$ with respect to the N-S outflow system of 4A1, suggesting a possible jet precession.

\smallskip
\noindent
\textit{4A2 outflow systems:}

Our observations unravel the intricate details of the 4A2 outflow systems, where two distinct outflow systems (C2-red \& C2-blue, and C4-red \& C4-blue) emerge, as detailed in the previous sections. 
The first system (C2-red \& C2-blue) extends approximately 18$''$ both to the north and south of 4A2, though the eastern lobe of the southern outflow appears less distinct. 
This outflow system has a PA of $\sim$29$^{\circ}$. 
The cavity walls move at a velocity of $\sim$16\,km\,s$^{-1}$ in the red, and they are observed with the three molecular tracers: SiO, H$_2$CO, and HDCO. Notably, we observe a velocity gradient in the red-shifted cavity walls, with the west wall exhibiting a velocity of approximately 3--7\,km\,s$^{-1}$ red-shifted compared to the east wall.
This intriguing observation hints at a potential rotation of the cavity, adding a layer of complexity to the dynamics of the 4A2 outflow system. Rotating outflow cavities have been detected in other objects (e.g. HH\,212 \citealt{Codella2014}, VLA1623 \citealt{Ohashi2022}).
We expect that if the jet is launched by a magnetic centrifugal mechanism it should rotate consistently with the disk (e.g. HH\,212 \citealt{Lee2017}) and transfer the rotational velocity also to the cavities.


The second outflow system stemming from 4A2 (C4-red \& C4-blue) exhibits a greater extent, reaching almost 25$''$ both to the north and 23$''$ to the south, though with a less distinct boundary in the southern direction.
This system has a PA of $\sim$26$^{\circ}$. 
The cavity walls move at a velocity of approximately 10\,km\,s$^{-1}$ in SiO and H$_2$CO, while HDCO does not exhibit these features. The absence of features in HDCO could stem from a sensitivity issue (not connected though to the excitation of the line), particularly considering that C4-red is the oldest cavity. It is worth noting that HDCO might also be frozen back onto the ice, further complicating its detection. A comprehensive astrochemical modelling could provide better insights into this matter.

Consequently, the two cavities of 4A2 show a slight misalignment of about 3$^{\circ}$, indicating a possible precession or rotation of the ejection. Notably, the cavity of the second, more extended outflow system displays lower velocities, aligning with the concept of an older ejection event that has decelerated in the current epoch. 

As mentioned in Sect. \ref{subsec:overview}, we associated the SiO emission labelled "C4-w" in Fig. \ref{fig:sketch} with the cavity wall of the second outflow system rather than a bend induced by the interaction of the jet with an obstructing dense core \citep{Choi2005}. The SiO(1--0) emission observed by \cite{Choi2005} are similar to that of SiO (2--1) observed more recently by \cite{DeSimone2022}, and to compare with our results we use the latter for convenience.  

In Fig. \ref{sio-choi}, we illustrate the SiO map observed by \cite{DeSimone2020} alongside those from our study, overlaid with two solid ellipses the two ellipses representing the cavities of the aforementioned two outflow systems. 
While the second ellipse encompasses "C4-w", it does not reach the farthest SiO blob observed by \cite{Choi2005, DeSimone2022} at $\sim$35$''$ from 4A2, prompting a discussion on its nature. 

One can speculate that a third event gave rise to a third outflow system, whose bow-shock coincides with the farthest SiO blob observed at $\sim$35$''$ from 4A2. We represent it with a dashed ellipse in Fig. \ref{sio-choi}. 
This ellipse would be inclined by 31$^{\circ}$, slightly deviating from the orientation of the other two ellipses (also see "Hypothetical cavity" in Fig. \ref{fig:sketch}). 
The sequence from the most distant ellipse (corresponding to the oldest ejection event) to the closest (and youngest event) reads as 31$^{\circ}$, 26$^{\circ}$, and 29$^{\circ}$, suggesting once more a precession or rotation of the ejections. This hypothesis is further supported by the morphology of the blue outflow lobe of 4A2, which bears a notable resemblance to the red lobe of the HH\,46/47 outflow, displaying a V-shape and a subsequent bifurcation in the cavity lobe. 
As proposed by \cite{Arce2013}, this feature could result from multiple mass ejections that were not detected in our observations. 

It is worth noting that the SiO blob at $\sim$35$''$ from 4A2 observed by \cite{Choi2005, DeSimone2022} is $\sim$6$''$ out of our FoV and was not detected in our observations due to filtering.

\smallskip
\noindent
\textit{The comparison between 4A1 and 4A2 ejections}

Our data shows that each of the protostars had at least two different episodic events. 
However, it is worth noting that the outflow cavities of 4A2, particularly C4-red, shows a much larger extension and is detected at lower velocities.
This would suggest that 4A2 is older than 4A1, as previously suggested (e.g. \citealt{Choi2007, Choi2011b, Santangelo2015, DeSimone2020}).

\subsection{Comparison with previous studies}\label{subsec:disc-comparison}

\smallskip
\noindent
\textit{Outflowing motion}: 
As highlighted in the Introduction, extensive studies have focused on the outflows in the IRAS\,4A system. 
While the blue-shifted emission from the southern region has been more clearly discernible when it comes to associating the observed feature with one of the protostars, the interpretation of the emission to the north of the protostars has proven to be challenging. 
This difficulty arises from the complex overlap between components originating from 4A1 and 4A2, hindering a clear understanding of the underlying processes in this region.

In the context of the jets within the IRAS\,4A system, our observations have unveiled the previously identified northern jet originating from 4A1 \citep{Santangelo2015, Taquet2020}. 
The enhanced angular resolution of our data has provided a clearer view of this jet, allowing us to resolve intricate bullet structures within it. Notably, our observations have also introduced a new element to the understanding of the system: a SiO component associated with the primary jet of 4A2. This component also exhibits bullet-like structures. 

Regarding the outflowing motion to the north of the protostars, previous studies presented different hypotheses regarding its origin. 
Some proposed it solely from the 4A2 outflow \citep{Choi2005}, while others attributed it to the bent outflow of 4A1 \citep{Ching2016}. 
Alternatively, certain studies argued for the presence of two extended bent cavities—one from 4A1 and another from 4A2 \citep{Chuang2021}. 
Our observations suggest a scenario distinct from these interpretations, as summarised in Table \ref{tab:summary-components}. 
We contend that neither of the cavities is bent. 
In fact, the non-detection of the bow-shock Bow1 in the previous studies has led to uncertainties regarding the extent and orientation of the cavity associated with 4A1, where it was thought that it is extended up to $\sim$17$''$ and bent to the east. 
Our observations propose that the cavity is oriented N-S and does not extend farther than 10$''$ from 4A1. 
It is also somewhat encompassed within that of 4A2, at least in projection. 
This N-S outflow cavity of 4A1 has a blue-shifted counterpart to the south of the protostar, where a second cavity (C3-blue) is also detected. 
The latter is misaligned with the former by $\sim$12$^{\circ}$. 
The blueshifted cavities are both present in the SO maps of \cite{Chuang2021}, but only C3-blue was identified.

In 4A2, previous studies suggested the existence of an extended outflow with a PA of approximately 18.9$^{\circ}$, that changes direction around 20$''$ (i.e., directional variability) attributed to an interaction with an obstructing dense core \citep{Choi2005}. 
Recent higher-resolution observations have confirmed the presence of an outflow cavity with a PA ranging from 19$^{\circ}$ to 22$^{\circ}$\citep[e.g.][]{Ching2016, Chuang2021}.
However, \cite{Marvel2008} noted that the position angle of the water maser outflow in IRAS\,4A differs from the position angles associated with the larger-scale outflows originating from the two sources within the system. 
Our current data reveal the existence of a high-velocity outflow cavity originating from 4A2, aligning closely with the jet observed in our dataset with a PA of $\sim$29$^{\circ}$. 
Furthermore, we distinguish between LV and HV outflow cavities in 4A2, clarifying that what was previously thought to be “directional variability” of the 4A2 outflow (C4-w) is simply an extension of the LV component of the western cavity wall.

\smallskip
\noindent
\textit{Jet precession}: 
The misalignment between the outflow systems in 4A1 and those in 4A2, suggests that the jets of the two sources are actually precessing. 
To verify that the misalignment is attributed to a precession rather than the potential shifting position of protostars over time, we examined the presence of proper motions of this source. 
We found no significant proper motion, with the coordinates showing only minor fluctuations within the uncertainties since their initial determination \citep{Looney2000, Reipurth2002}, as compared to recent derivations \citep{desimone2020vla}. 
While the possibility of precession in the 4A1 jet has not been proposed before, previous studies hinted at the precession scenario in 4A2 to explain specific features observed in the IRAS\,4A system (e.g. \citealt{Choi2006, Santangelo2015, Chuang2021}). 
\cite{Choi2006} referred to the mechanism as “drifting” rather than precession due to a lack of evidence for periodic change. 
In another work, \cite{Choi2011a} have suggested that the 4A2 jet is rotating within a cone of opening half-angle $\sim$3.1$^{\circ}$. 
The cause of the 4A2 jet's precession remains unclear. 
Several studies suggested that the precession might result from a close unresolved binary system, with a separation of 30-80\,au \citep{Choi2006}, 32\,au \citep{Marvel2008} or larger than $\sim$12\,au \citep{Chuang2021}. 
Alternatively, \cite{Chuang2021} proposed that a misalignment between the magnetic field and the cloud's angular momentum vector may induce precession. 
This scenario may also apply to IRAS\,4A1. Particularly noteworthy is the resemblance of the blueshifted outflow cavities to the misaligned outflows observed in HOPS 373, where one cavity is “nested” within the other \citep{Lee2024}. The authors propose that the observed misaligned outflows may originate from either a single protostar, following a reorientation of its angular momentum axis, or from a close protostellar binary, with each outflow linked to a distinct binary component. They further suggest that, in the first scenario, the reorientation could result from a change in the angular momentum axis of the accreting gas or from a misalignment between the rotation axis of the core and the orientation of the global magnetic field (e.g. \citealt{Machida2020, Okoda2021}). 
With our observations, it is challenging to determine the origin of the precession of 4A1 and 4A2' jets, as the identification of a potential binary companion for each protostar would require an angular resolution of about 12\,au. A reorientation of the system could be plausible given the misalignment of the magnetic field in the IRAS 4A envelope with the rotation axis of the circumbinary envelope (\citealt{Chuang2021} and references therein). Further observations with higher angular resolution are needed to resolve the discs and investigate the possibility of close binaries for each protostar.


\smallskip
\noindent

\textit{Effect of stellar multiplicity}: The influence of stellar multiplicity on outflow activity during the protostellar stage remains inadequately characterised. One notable effect is the precession of the disc, which can alter the orientation and morphology of the outflows. Precession refers to the rotation of the jet axis within a cone, leading to changes in the direction of the outflow. This process was observed towards several sources and was linked to the presence of a binary companion (e.g. HOPS-370 \citealt{Tobin2019-orion, Chahine2022b}), or a close unresolved binary (e.g. L1157 \citealt{Podio2016}, CepE-A \citealt{Schutzer2022}, HOPS-317 in the HH\,24 complex \citealt{Reipurth2023}, HOPS-373SW \citealt{Lee2024})
Additionally, binary and multiple systems may exhibit jet wiggling, where the orientation of the jet remains constant, but the spatial position of the launching point varies as the stars orbit each other (e.g. HH\,211 \citealt{Lee2010}, HH\,212 \citealt{Lee2017}, CepE-A \citealt{Schutzer2022}, VLA15 in OMC-2\,FIR\,4 \citealt{Chahine2022b}, SSV63 and Jet C in the HH\,24 complex \citealt{Reipurth2023}). Furthermore, the disturbance of the disc due to the presence of a companion can lead to increased accretion onto the protostars, potentially driving strong outbursts of outflow activity (e.g. HH\,175 \citealt{Reipurth2021}). While precession has been observed in the IRAS\,4A system, potentially linked to stellar multiplicity, further investigations are warranted, particularly at the scale of the circumstellar discs. This additional scrutiny is necessary to discern any signs of disc disturbance or misaligned accretion onto the protostars.

%% file: sections/conclusion.tex
\label{sec:conclusions}

In this study, we delved into the morphology and kinematics of the jets and outflows towards the IRAS\,4A system using three distinct tracers: SiO, H$_{2}$CO, and HDCO, at a spatial resolution of $\sim$150 au. The use of these three tracers proved to be indispensable in revealing new features, piecing together the intricate puzzle, and gaining a comprehensive understanding of the larger framework, as each tracer contributed a crucial component to the analysis. Our main conclusions are summarised below.

\begin{enumerate}

    \item We have successfully identified two distinct jets in the redshifted emission, originating from 4A1 and 4A2. The jet from 4A1 is oriented in a N-S direction, while that from 4A2, observed for the first time, exhibits a NE-SW orientation with a PA of 29$^{\circ}$.
    \item Each protostar has two outflow systems, with 4A1 showcasing three outflow cavities and 4A2 presenting four. Contrary to prior studies, we argue that none of the outflows is bent.
    \item We observe an outflow cavity to the north of IRAS\,4A1 and two cavities to its south. One of these southern cavities aligns with a N-S direction, while the other exhibits a PA of -12$^{\circ}$ without a corresponding counterpart in the north. The misalignment of these cavities suggests the possibility of a precession occurring within the 4A1 jet.
    \item The analysis of the outflows originating from IRAS\,4A2 reveals the presence of two cavities in both the northern and southern directions, albeit with less distinct emission in the south. The cavities are slightly misaligned, and the more extended is moving at lower velocities, suggesting it is potentially stemming from an older ejection event. In addition, the cavity moving at higher velocities shows an indication of rotation
    \item Our study challenges previous interpretations of the directional variability observed in the outflow of IRAS\,4A2, suggesting instead that it represents the remnants of the western wall of the low-velocity outflow cavity. We speculate on the occurrence of a third event that was not detected in our data.
\end{enumerate}
In summary, our study underscores the importance of utilising diverse tracers beyond traditional ones, emphasising the significance of comprehensive studies in unravelling the complexities of outflow systems.

%% file: sections/apendix.tex
\label{appendix}

\begin{figure}

    \centering
    \includegraphics[width=0.42\textwidth]{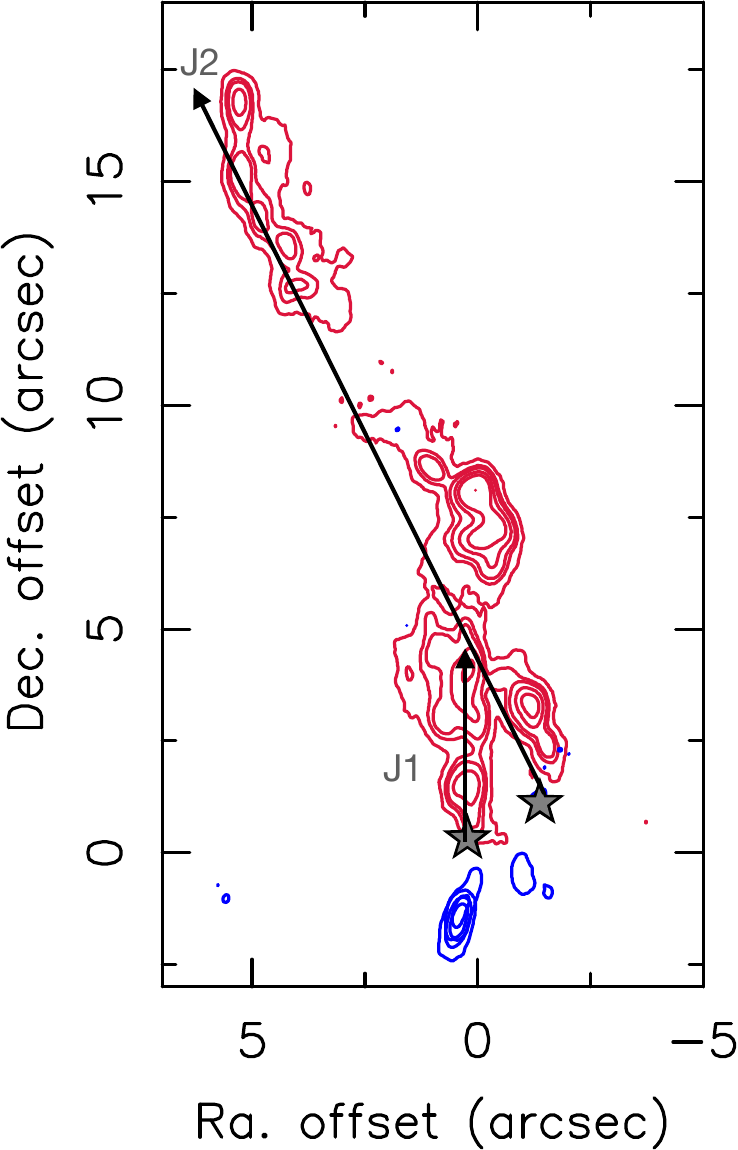} 

    \caption{Zoomed view of the SiO emission within the HV1 velocity interval. The contours are as in the HV1 panel of Fig. \ref{sio-mom-chan}. The positions of IRAS 4A1 and IRAS 4A2 are marked with stars. The jets from each source are marked with arrows.}
    
   \label{sio-jets}
\end{figure}

\begin{figure*}
    \centering
    \includegraphics[width=0.98\textwidth]{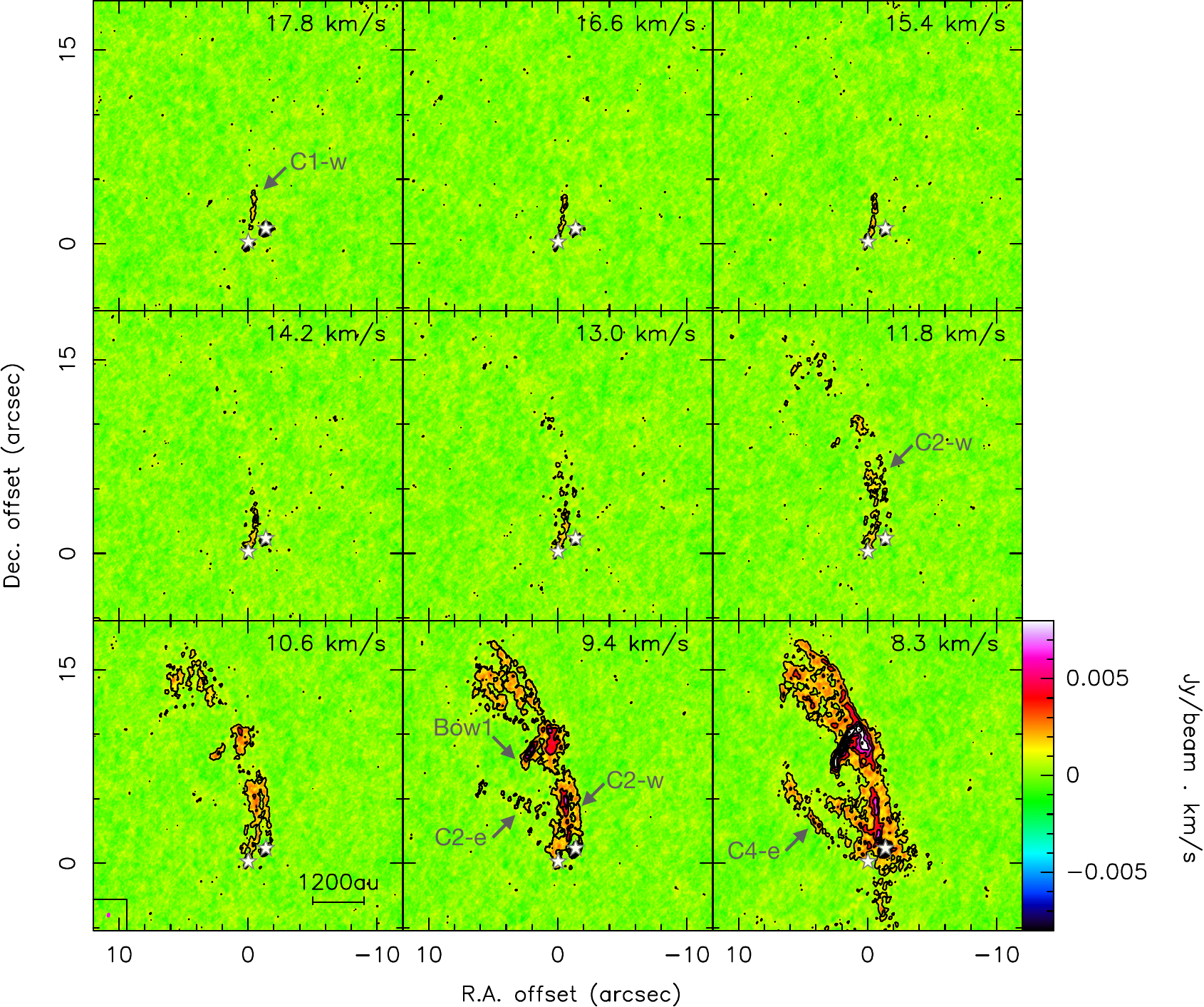} 

    \caption{Channel map of the redshifted emission of HDCO. The contour levels start at 3$\sigma$ and increase by 3$\sigma$ step with $\sigma$\,=\,0.40\,mJy\,beam$^{-1}$\,km\,s$^{-1}$. The velocity of each channel is indicated in the upper-right corner of each panel. The features mentioned in the text are labelled in grey. The synthesised beam is depicted in the lower-left corner of the bottom-left panel.}
    
   \label{hdco-chan}
\end{figure*}

\begin{figure*}
    \centering
    \includegraphics[width=0.98\textwidth]{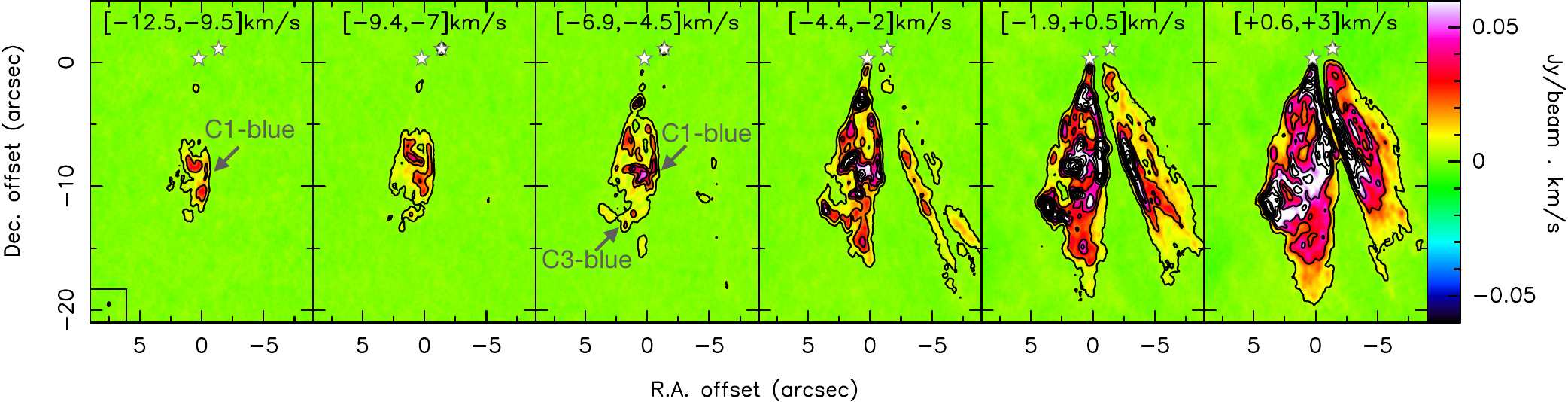} 

    \caption{Channel map of the blueshifted emission of H$_{2}$CO between -12.5 and +3\,km\,s$^{-1}$. The velocity intervals and their corresponding labels are shown at the upper right corners of each channel map panel. The contour levels start at 5$\sigma$ and increase by 20$\sigma$ step with $\sigma$\,=\,0.75\,mJy\,beam$^{-1}$\,km\,s$^{-1}$. The features mentioned in the text are labelled in grey. The positions of IRAS\,4A1 and IRAS\,4A2 are marked with stars. The synthesised beam is depicted in the lower-left corner of the bottom-left panel.}
    
   \label{h2co-chan-blue}
\end{figure*}

\begin{figure*}
    \centering
    \includegraphics[width=0.98\textwidth]{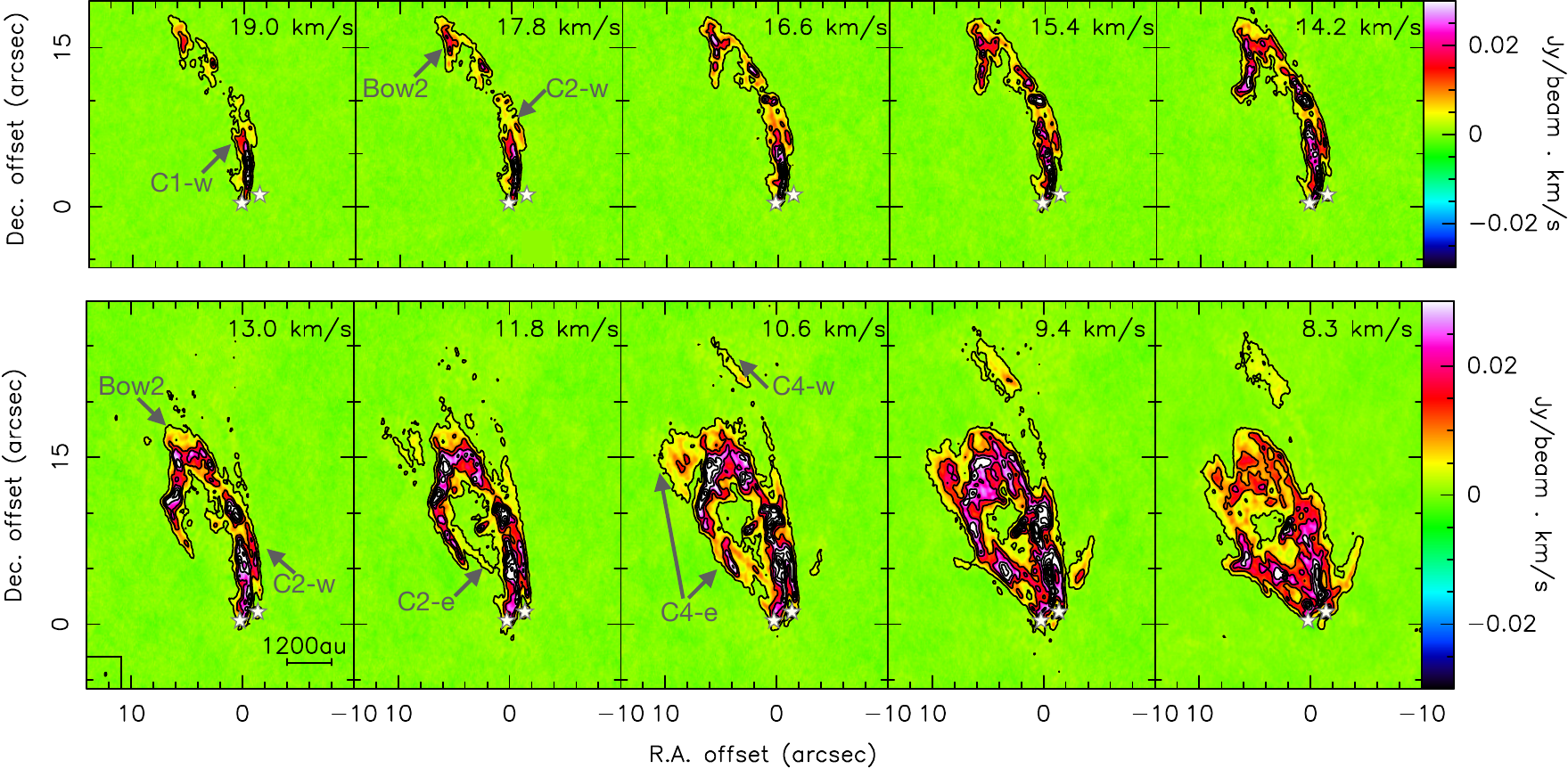} 

    \caption{Channel map of the redshifted IV and LV emission of H$_{2}$CO. Each panel displays the integrated velocity range of $\pm$0.6\,km\,s$^{-1}$ centred around the velocity specified in the upper right corner. The contour levels start at 5$\sigma$ and increase by 20$\sigma$ step with $\sigma$\,=\,0.45\,mJy\,beam$^{-1}$\,km\,s$^{-1}$. The features mentioned in the text are labelled in grey. The positions of IRAS\,4A1 and IRAS\,4A2 are marked with stars. The synthesised beam is depicted in the lower-left corner of the bottom-left panel.}
    
   \label{h2co-chan-red}
\end{figure*}

\begin{figure*}
    \centering
    \includegraphics[width=0.33\textwidth]{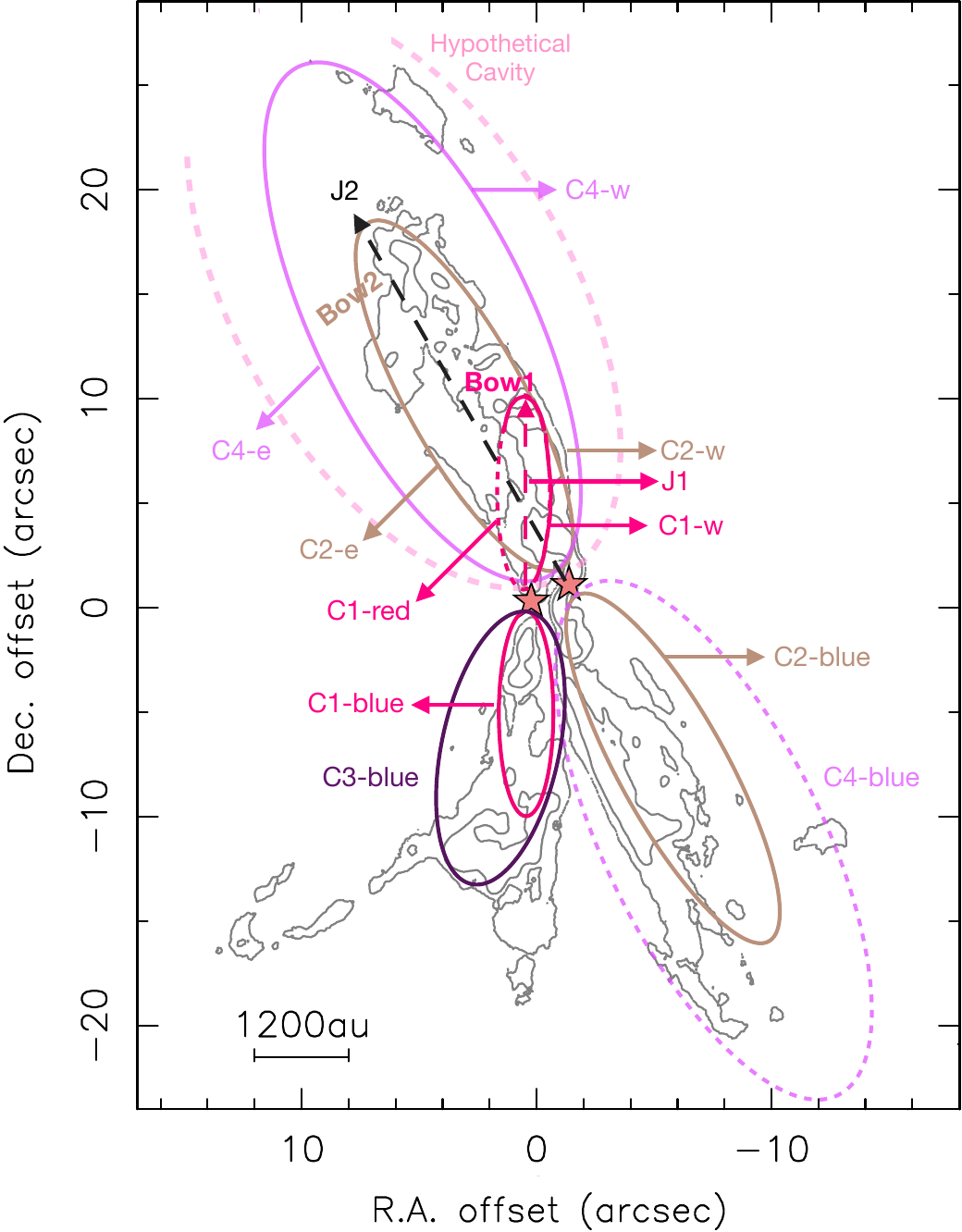} 
     \includegraphics[width=0.33\textwidth]{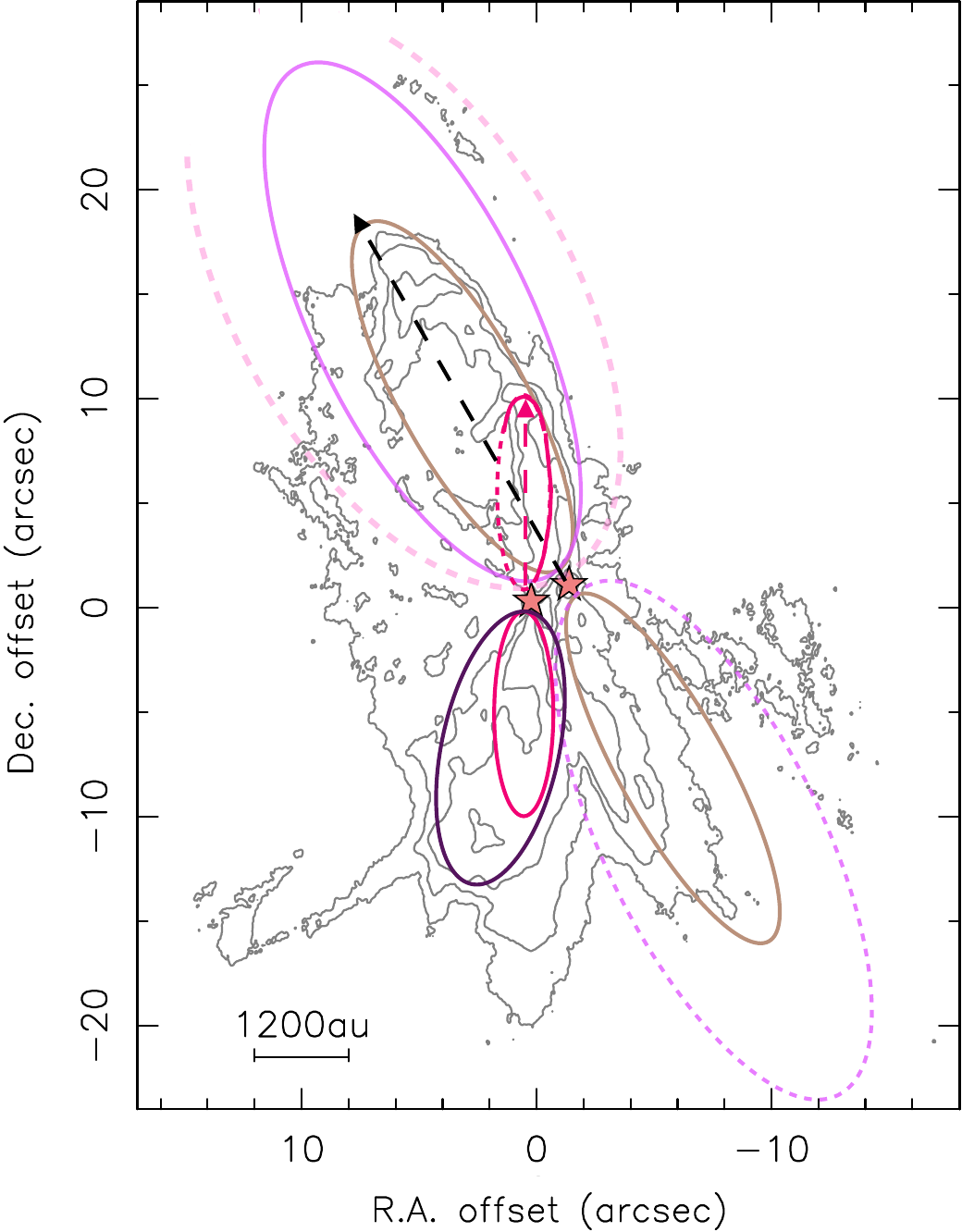}
     \includegraphics[width=0.33\textwidth]{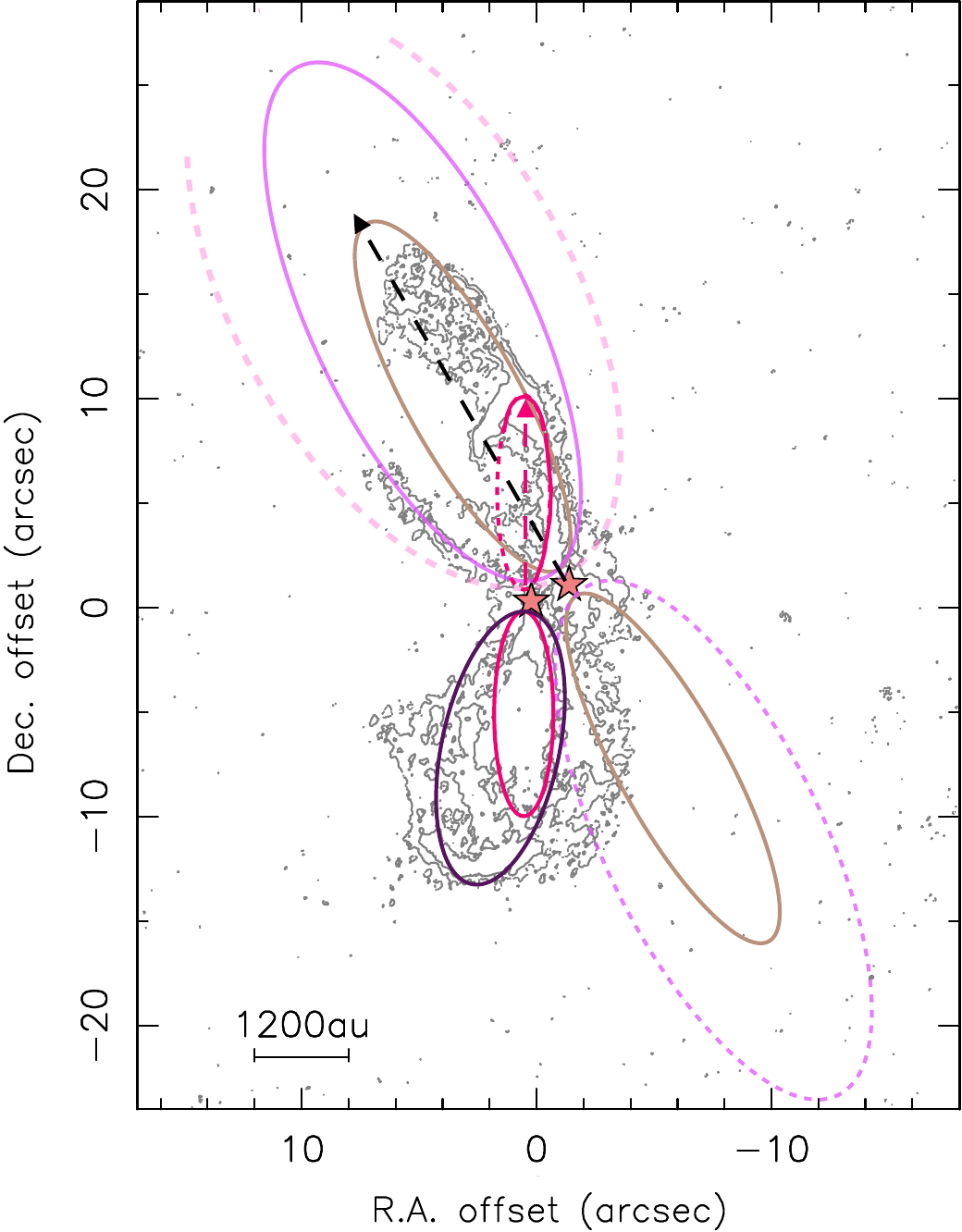}
    \caption{Integrated intensity maps of SiO, H$_{2}$CO and HDCO towards IRAS\,4A with overlaid ellipses and arrows representing all the features identified in this work. The velocity range are as in the main text. The contours of SiO are at 5$\sigma$ and 150$\sigma$ ($\sigma$\,=\,0.81\,mJy\,beam$^{-1}$\,km\,s$^{-1}$), those of H$_{2}$CO are at 5, 20$\sigma$ and 40$\sigma$ ($\sigma$\,=\,5\,mJy\,beam$^{-1}$\,km\,s$^{-1}$) and the ones of HDCO are at 5$\sigma$ and 25$\sigma$ ($\sigma$\,=\,0.4\,mJy\,beam$^{-1}$\,km\,s$^{-1}$). The positions of IRAS\,4A1 and IRAS\,4A2 are marked with stars. All the feature identified in this work and summarised in both Fig. \ref{fig:sketch} and Table. \ref{tab:summary-components} are represented with ellipses and arrows.}
    
   \label{all-mom-ellipses}
\end{figure*}

In Fig. \ref{sio-jets} we show a zoomed view of the SiO emission within the HV1 velocity interval.
In Figs. \ref{hdco-chan} and \ref{h2co-chan-red} we report the channels maps of HDCO and H$_{2}$CO in the redshifted LV and IV regimes. In Fig. \ref{h2co-chan-blue} we present the detailed blueshifted channel map of H$_{2}$CO.
In Fig. \ref{all-mom-ellipses} we show the integrated intensity maps of the different tracers used to probe the emission towards IRAS\,4A, with superimposed ellipses and arrows representing all the features identified in this work.



    